\documentclass[preprint,11pt]{elsarticle}
\usepackage[T1]{fontenc}
\usepackage[latin9]{inputenc}
\usepackage{units}
\usepackage{amsmath}
\usepackage{amssymb}
\usepackage{graphicx,epstopdf}
\usepackage{esint}
\usepackage{color}

\journal{The Journal of Non-Newtonian Fluid Mechanics}

\makeatletter

\usepackage{subfig}
\makeatother

\begin{document}

\begin{frontmatter}

\title{Shear-thinning in dense colloidal suspensions and its effect on
elastic instabilities: from the microscopic equations of motion to an
approximation of the macroscopic rheology}

\author[AN1,AN2]{Alexandre NICOLAS\corref{ANaddress}}
\author[MF]{Matthias FUCHS}

\address[AN1]{Universit\'e Grenoble-Alpes, LIPhy, F-38000 Grenoble, France}
\address[AN2]{CNRS, LIPhy, F-38000 Grenoble, France}

\address[MF]{Fachbereich Physik, Universit\"at Konstanz, 78457 Konstanz,
Germany}

\cortext[ANaddress]{Corresponding author's email:
alexandre.nicolas@polytechnique.edu; \\Phone: (+54)9.2944.81.9476.\\
Present address:
CONICET \& Centro At\'omico Bariloche, Bariloche, Argentina.}

\begin{abstract}
In the vicinity of their glass transition, dense colloidal suspensions
acquire elastic properties over experimental timescales. We investigate
the possibility of a visco-elastic flow instability in curved geometry
for such materials. To this end, we first present a general strategy extending
a first-principles approach based on projections onto slow variables
(so far restricted to strictly homogeneous flow) in order to
handle inhomogeneities. In particular, we separate the advection of the
microstructure by the flow, at the origin of a fluctuation advection term, from
the intrinsic dynamics. On
account of the complexity of the involved equations, we then opt for
a drastic simplification of the theory, in order to establish its potential to
describe instabilities.  
These very strong approximations lead to a constitutive
equation of the White-Metzner class, whose parameters are fitted with
experimental measurements of the
macroscopic rheology of a glass-forming colloidal dispersion. The model
properly accounts for the shear-thinning properties of the dispersions, but,
owing to the approximations, the description is not fully quantitative. Finally,
we perform a linear stability analysis of the flow in the experimentally
relevant cylindrical (Taylor-Couette) geometry and provide evidence that
shear-thinning strongly
stabilises the flow, which can explain why visco-elastic instabilities
are not observed in dense colloidal suspensions.
\end{abstract}


\end{frontmatter}

\section{Introduction}

\subsection{Observations}

Take a small amount of carbon black powder and disperse it into water:
this gives pigmented ink. From a rheological perspective, it is a
colloidal suspension which flows similarly to water, albeit with a
somewhat higher viscosity. But this Newtonian behaviour, which holds
generically for very dilute suspensions, is strongly altered when
the volume fraction $\phi$ of colloids gets larger. Most strikingly,
the viscosity and relaxation time of the suspension increase dramatically
when $\phi$ approaches a {}``critical'' packing fraction $\phi_{g}$
($\phi_{g}\approx0.56$ for hard-sphere-like colloids \cite{Pusey1987}).
For $\phi\gtrsim\phi_{g}$, the material retains elastic properties
over any experimental timescale, in a fashion reminiscent of the emergence
of glassiness in supercooled melts of some metallic alloys, when the
temperature declines \cite{Amann2013}. Despite these dramatic changes,
the structure of the material remains essentially liquid-like throughout
the transition. Accordingly, it appears sensible to compare the rheology
of very dense colloidal suspensions to that of other visco-elastic
liquids. In particular, one may wonder why a variety of complex fluids
among the latter, such as worm-like micelles or polymer solutions
\cite{Muller1989,Fardin2012a}, are prone to a (non-inertial) flow
instability in curved geometry, leading for instance to the formation
of vortices, while, to the best of our knowledge, no such visco-elastic
instability has ever been reported in very dense colloidal suspensions.

\subsection{A microscopic approach using mode-coupling theory}

The level of difficulty required to rationalise the rheology of suspensions
strongly depends on the volume fraction $\phi$ of interest. In the
dilute regime, the fluid is Newtonian; its viscosity is independent
of the applied shear rate. More quantitatively, the linear corrections
to the solvent viscosity due to the colloids were worked out by Einstein
a little more than a century ago, under the assumption of non-interacting
colloids \cite{Einstein1906}. By a detailed study of the probability distribution
function of particle pairs, the approach was extended to interacting colloids by  Batchelor
and others \cite{Batchelor1977effect}, and led to a description of the semi-dilute regime \cite{Brady1996model}.   
For $\phi\approx\phi_{g}$, collective effects become paramount, in that glassiness
can be thought of as the entrapment of particles in the {}``cages''
formed by their neighbours; these effects turn a first-principles
derivation of the macroscopic rheology into a formidable challenge,
all the more so as the presence of flow distorts the structure of
the material away from its {}``quiescent configuration'' and gives
rise to complex interplays \cite{Brader2010nonlinear}.

Nevertheless, at the expense of some uncontrolled approximations,
the mode-coupling theory developed by Sj\"ogren, Bengtzelius, G\"otze,
Sjolander, and others \cite{Bengtzelius1984,Sjogren1980} succeeded
in rationalising the phenomenology of the glass transition by focusing
on the evolution and the relaxation of the (slow) density modes of
the system and on their coupling to the other (faster) variables.
In the last decade, Fuchs, Cates \emph{et al.} \cite{Fuchs2002,Brader2012},
and Miyazaki \emph{et al.} \cite{Miyazaki2004} in a parallel endeavour,
were able to extend this framework to situations of flow, in which
the colloids are dragged by a prescribed solvent flow. The state
of the art of this theory encompasses arbitrary, potentially time-dependent
incompressible solvent flows, in two or three dimensions 
\cite{Brader2012,Amann2015nonlinear}.

However, the derivation hinges on the assumption of a perfectly homogeneous
flow throughout space; this hampers the investigation of any
flow instability. Indeed, perturbations, which break homogeneity,
are not handled adequately; in particular, the mechanism describing
their (expected) advection with the flow is still missing in the equations.
Moreover, the complexity of the final equations giving the stress as a function of the
strain history is a deterrent to any stability analysis in non-trivial geometry.

\subsection{Objectives of the article}

In this contribution, we first propose a general way to extend the formalism
and handle flow inhomogeneities,
insisting in particular on the recovery of a fluctuation advection
term and on the limit of locally homogeneous flow.
Then, we follow the endeavour pioneered in Ref.~\cite{Brader2009}
to reduce the final equations to a tractable constitutive equation.
This will come at the expense of very strong (but explicitly
exposed) approximations and clearly undermine the accuracy of the
description. Nevertheless, the ensuing simple model, which falls in the White-Metzner
class \cite{White1963}, will allow us to capture the
experimentally measured low-shear-rate rheology and high-shear-rate
rheology in a model colloidal glass-forming dispersion \cite{Siebenbuerger2012,Voigtmann2014}. Finally, a linear stability analysis of the flow will be
performed, in cylindrical (Taylor-Couette) geometry and the (stabilising) effect of effect of
shear-thinning on the visco-elastic flow will be numerically assessed.

\section{Theoretical probabilistic framework\label{sec:Theoretical-framework-for}}

We start by presenting the theoretical underpinning of the rheological
equations that extend quiescent mode-coupling theory. 

Let us consider an assembly of $N$ colloidal particles dispersed
in a solvent and evolving by Brownian motion in a volume $V$, for instance with periodic
boundary conditions.

\subsection{From the overdamped Langevin equation to the Smoluchowski equation}

To describe the microscopic motion of particle $i\in\left\{ 1,\ldots,N\right\} $,
we posit an overdamped Langevin equation acting on its velocity $\boldsymbol{\dot{r}_{i}}$:

\begin{equation}
\zeta\left[\boldsymbol{\dot{r}_{i}}-\boldsymbol{v}^{\mathrm{solv}}\left(\boldsymbol{r_{i}}\right)\right]=\boldsymbol{F_{i}}+\boldsymbol{f_{i}}^{\mathrm{th}}\label{eq:overdamped_Langevin}
\end{equation}
Here, $\boldsymbol{F_{i}}$ is the conservative force that derives
from the global potential energy of the system, and the $\boldsymbol{f_{i}}^{\mathrm{th}}$'s
are random Gaussian thermal fluctuations, \emph{viz.}, $\left\langle \boldsymbol{f_{i}}^{\mathrm{th}}\left(t\right)\right\rangle =\boldsymbol{0}$
and $\left\langle \boldsymbol{f_{i}}^{\mathrm{th}}\left(t\right)\otimes\boldsymbol{f_{j}}^{\mathrm{th}}\left(t^{\prime}\right)\right\rangle =2k_{B}T\zeta\delta_{ij}\delta\left(t-t^{\prime}\right)\mathbb{I}$,
where $\mathbb{I}$ is the identity matrix in \emph{$d$ }dimensions.
The frictional force on the left-hand side (lhs) involves a frictional coefficient
$\zeta$ and the particle velocity relative to the (prescribed) local
solvent velocity $\boldsymbol{v}^{\mathrm{solv}}\left(\boldsymbol{r}\right)$;
for an incompressible flow, the latter should satisfy $\nabla\cdot\boldsymbol{v}^{\mathrm{solv}}=0$.
Hydrodynamic effects, being presumably subordinate to short-range
interactions in dense systems, are neglected.

Rather than focusing on the motion of individual particles, we adopt
a statistical approach. Equation \ref{eq:overdamped_Langevin} is
recast into the following equations for the evolution of the probability
$\psi\left(\Gamma;t\right)$ to find the system in the microscopic
configuration $\Gamma\equiv\left(\boldsymbol{r_{1}},\ldots,\boldsymbol{r_{N}}\right)$
at time $t$ \cite{Risken1989}:
\begin{equation}
\begin{cases}
\psi(\Gamma;t=0) & =\psi_{0}(\Gamma)\\
\partial_{t}\psi(\Gamma;t) & =\Omega(\Gamma;t)\psi(\Gamma;t).
\end{cases}\label{eq:Smoluchowski1}
\end{equation}
Time evolution is given by  the Smoluchowski operator

\[\Omega(\Gamma;t)\equiv\sum_{i=1}^{N}\partial_{i}\cdot\left[\partial_{i}
-\boldsymbol{F_{i}}\left(\Gamma\right)-\boldsymbol{v}^{\mathrm{solv}}
\left(\boldsymbol{r_{i}},t\right)\right],
\]
where $\partial_{i}\equiv\frac{\partial}{\partial\boldsymbol{r_{i}}}$
and we have used dimensionless units by setting $\zeta=1$ and $k_BT=1$.  Here,
contrary
to Ref.~\cite{Fuchs2002,Fuchs2009,Brader2009,Brader2012},
the initial probability density $\psi_{0}(\Gamma)$ need not be the
equilibrium distribution: the system can be prepared
in an arbitrary configuration.

Equation~\ref{eq:Smoluchowski1} is formally solved by
\begin{equation}
\psi(\Gamma;t)=e_+^{\int_{0}^{t}\Omega(\Gamma;s)ds}\psi_{0}(\Gamma),\label{eq:formal_solution_psi}
\end{equation}
where $e_+$ is a time-ordered product (see Appendix A of Ref.~\cite{Brader2012}).
At time $t$, the evaluation of a many-body function $g$ reads
\begin{equation}
\left\langle g\right\rangle _{t}\equiv\int g\left(\Gamma\right)\psi(\Gamma;t)d\Gamma\label{eq:g_t}
\end{equation}

Instead of having the probability distribution $\psi$ evolve in time,
as in Eq.~\ref{eq:formal_solution_psi}, a dual formulation is sometimes
preferable, in which (by means of a partial integration of Eq.~\ref{eq:g_t})
$\psi$ is kept constant and the definiton of $g$
evolves with time, analogously to the switch from a wavefunction-evolving
Schr\"odinger representation to an operator-evolving Heisenberg representation
in Quantum Mechanics, \emph{viz.},
\begin{equation}
\begin{cases}
\psi(\Gamma;t) & =\psi_{0}(\Gamma)\\
\partial_{t}g(\Gamma;t) & =\Omega^{\dagger}(\Gamma;t)g(\Gamma;t),
\end{cases}\label{eq:adjoint_Smoluchowski}
\end{equation}
where $\Omega^{\dagger}(\Gamma;t)\equiv\sum_{i=1}^{N}\left[\partial_{i}+\boldsymbol{F_{i}}\left(\Gamma\right)+\boldsymbol{v}^{\mathrm{solv}}\left(\boldsymbol{r_{i}},t\right)\right]\cdot\partial_{i}$
is the adjoint of the Smoluchowski operator, with the formal solution
\begin{equation}
g(\Gamma;t)=e_{-}^{\int_{0}^{t}\Omega^{\dagger}(\Gamma;s)ds}g(\Gamma;0),\label{eq:g_t_formal_sol}
\end{equation}
where $e_-$ denotes the negatively ordered exponential \cite{Brader2012}.

\subsection{Auxiliary frame and recovery of an advection
term\label{sub:Advection_term_gal}}

Microscopic observables depend on space, \emph{via} their point of
evaluation $\boldsymbol{r}$: $g(\Gamma;t)\rightarrow g(\boldsymbol{r};\Gamma;t)$.
But the prescribed velocity field generally differs from zero at $\boldsymbol{r}$,
so that the evolution of $g(\boldsymbol{r})$ mingles an intrinsic
evolution of the system and an advection by the flow field. In
previous studies, for instance, Ref.~\cite{Brader2012}, the consideration
of a strictly homogeneous system (with vanishing spatial gradients) rendered a disentanglement of the
two effects unnecessary and no advection term appeared in the equations. Yet,
in the presence of any heterogeneity, such term is expected on physical
grounds and is crucial for the study of perturbations, hence, instabilities.
Here, we purport to carefully establish its recovery.

To disentangle advection and intrinsic dynamics, it is helpful to
observe the dynamics in a frame that moves with the solvent velocity
at the point $\boldsymbol{r}_{o}$ and time $t_{o}$ that will be
of interest. Thus, we introduce new, time-dependent coordinates
\begin{equation}
\boldsymbol{r}^{\prime}\left[\boldsymbol{r},t\right]\equiv\boldsymbol{r}-\left(\boldsymbol{r}_{o}(t)-\boldsymbol{r}_{o}\right),\label{eq:change_of_coord}
\end{equation}
with the backward transform

\[
\boldsymbol{r}\left[\boldsymbol{r}^{\prime},t\right]=\boldsymbol{r}^{\prime}+\left(\boldsymbol{r}_{o}(t)-\boldsymbol{r}_{o}\right),
\]
where $\boldsymbol{r}_{o}(t)$ is the pathline of the (non-singular)
solvent velocity field
$\boldsymbol{v}^{\mathrm{solv}}\left(\boldsymbol{r},t\right)$
that ends at $\boldsymbol{r}_{o}$ at time $t_{o}$, \emph{i.e.}, 
\begin{equation}
\begin{cases}
\partial_{t}\boldsymbol{r}_{o}(t) & =\boldsymbol{v}^{\mathrm{solv}}\left(\boldsymbol{r}_{o}(t),t\right)\\
\boldsymbol{r}_{o}(t_{o}) & =\boldsymbol{r}_{o}.
\end{cases}
\label{eq:change_of_coordinates}
\end{equation}
(In ~\ref{sec:Operator-formulation-of}, we propose an equivalent,
alternative approach, rooted in operator formalism rather than change
of frame). At a fixed point $\boldsymbol{r}^{\prime}$ in the new
frame, $g$ evolves with time as follows:

\begin{align}
D_{t}g\left(\boldsymbol{r}\left[\boldsymbol{r}^{\prime},t\right],\Gamma;t\right) 
& \equiv
\underset{dt\rightarrow0}{\mathrm{lim}}\frac{g\left(\boldsymbol{r}\left[
\boldsymbol{r}^{
\prime},t+dt\right],\Gamma;t+dt\right)-g\left(\boldsymbol{r}\left[\boldsymbol{r}
^{\prime},t\right],\Gamma;t\right)}{dt} \label{eq:dt_g_aux}
\\ \nonumber
 & =  \Omega^{\dagger}(\Gamma;t)g\left(\boldsymbol{r}\left[\boldsymbol{r}^{\prime},t\right],\Gamma;t\right)+\partial_{t}\boldsymbol{r}_{o}(t)\cdot\partial_{\boldsymbol{r}}g\left(\boldsymbol{r}\left[\boldsymbol{r}^{\prime},t\right],\Gamma;t\right)
 \\ \nonumber
 & =
\Omega^{\dagger}(\Gamma;t)g\left(\boldsymbol{r}\left[\boldsymbol{r}^{\prime},
t\right],\Gamma;t\right)+\boldsymbol{v}^{\mathrm{solv}}\left(\boldsymbol{r}_{o}
(t),t\right)\cdot\partial_{\boldsymbol{r}}g\left(\boldsymbol{r}\left[\boldsymbol
{r}^{\prime},t\right],\Gamma;t\right).
\end{align}

Next, we notice that commonly used observables, such as the stress
or the density, do not depend \emph{intrinsically} on space, \emph{i.e.},
there exists a function $\tilde{g}$ such that $g(\boldsymbol{r},\Gamma)\equiv g(\boldsymbol{r},\boldsymbol{r_{1}},\ldots,\boldsymbol{r_{N}})=\tilde{g}(\boldsymbol{r_{1}}-\boldsymbol{r},\ldots,\boldsymbol{r_{N}}-\boldsymbol{r})$.
Consequently,
\begin{eqnarray}
\partial_{\boldsymbol{r}}g(\boldsymbol{r},\Gamma) & = & \partial_{\boldsymbol{r}}\tilde{g}(\boldsymbol{r_{1}}-\boldsymbol{r},\ldots,\boldsymbol{r_{N}}-\boldsymbol{r}) 
\\
 & = & -\sum_{i}\partial_{i}\tilde{g}(\boldsymbol{r_{1}}-\boldsymbol{r},\ldots,\boldsymbol{r_{N}}-\boldsymbol{r})\nonumber \\
 & = & -\sum_{i}\partial_{i}g(\boldsymbol{r},\Gamma).\label{eq:d_r_EQUALS_sum_di}
\end{eqnarray}

Inserting this result into Eq.~\ref{eq:dt_g_aux}, we get
\begin{eqnarray*}
& D_{t}g\left(\boldsymbol{r}\left[\boldsymbol{r}^{\prime},t\right],\Gamma;t\right) 
 =  \left[
\sum_{i=1}^{N}\left[\partial_{i}+\boldsymbol{F_{i}}
\left(\Gamma\right)+\boldsymbol{v}^{\mathrm{solv}}\left(\boldsymbol{r_{i}},
t\right)\right]\cdot\partial_{i}\right]
g\left(\boldsymbol{r}\left[\boldsymbol{r}^{\prime},t\right],\Gamma;t\right) &
\\
&  \:\:\:\:-\boldsymbol{v}^{\mathrm{solv}}\left(\boldsymbol{r}_{o}(t),t\right)\cdot\sum_{i}\partial_{i}g\left(\boldsymbol{r}\left[\boldsymbol{r}^{\prime},t\right],\Gamma;t\right)&
\\
 & =  \left[
\sum_{i=1}^{N}\left[\partial_{i}+\boldsymbol{F_{i}}
\left(\Gamma\right)+\left(\boldsymbol{v}^{\mathrm{solv}}\left(\boldsymbol{r_{i}}
,t\right)-\boldsymbol{v}^{\mathrm{solv}}\left(\boldsymbol{r}_{o}(t),
t\right)\right)\right]\cdot\partial_{i}\right]
g\left(\boldsymbol{r}\left[\boldsymbol{r}^{\prime},t\right],\Gamma;t\right) &
\end{eqnarray*}

Denoting by a prime the functions expressed in the new frame, \emph{i.e.},
$f^{\prime}\left(\boldsymbol{r}^{\prime}\left[\boldsymbol{r},t\right],t\right)=f\left(\boldsymbol{r},t\right)$
for a generic function $f$, and remarking that coordinates in the original and new frame are in a one-to-one correspondence, we arrive at 

\begin{eqnarray}
D_{t}g\left(\boldsymbol{r}\left[\boldsymbol{r}^{\prime},t\right],
\Gamma;t\right) &
\equiv & \partial_{t}g^{\prime}\left(\boldsymbol{r}^{\prime},\Gamma^{
\prime};t\right) 
 \notag
\\
& = &
\Omega^{\dagger\,\prime}(\Gamma^{\prime},t)g^{\prime}\left(\boldsymbol{r}^{
\prime},\Gamma;t\right),\label{dt_g_prime}
\end{eqnarray}
for any $\boldsymbol{r}^{\prime}$ in the domain, where 
\[
\Omega^{\dagger\,\prime}(\Gamma^{\prime},t)\equiv\sum_{i=1}^{N}\left[\partial_{i
}^{\prime}+\boldsymbol{F_{i}}^{\prime}\left(\Gamma^{\prime}\right)+\boldsymbol{
v}^{\prime}\left(\boldsymbol{r_{i}}^{\prime},t\right)\right]
\cdot\partial_{i}^{\prime}
\]
and
\begin{equation}
\boldsymbol{v}^{\prime}\left(\boldsymbol{r}^{\prime},t\right)\equiv\boldsymbol{v
}^{\mathrm{solv}\,\prime}\left(\boldsymbol{r}^{\prime},t\right)-\boldsymbol{v}^{
\mathrm{solv}\,\prime}\left(\boldsymbol{r}_{o},t\right).\label{eq:v_tilde_prime}
\end{equation}
Thus, an observable $g^{\prime}$ evaluated at fixed position in the
auxiliary frame displays dynamics identical to those of its
counterpart $g$ in the original frame (see Eq.~\ref{eq:adjoint_Smoluchowski}),
except that the velocity field $\boldsymbol{v}^{\mathrm{solv}}$ entering
the Smoluchowski operator for $g$ is replaced by a new field
$\boldsymbol{v}^{\prime}$
for $g^{\prime}$, which vanishes at $\boldsymbol{r}_{o}$.  Using Eq.~\ref{eq:v_tilde_prime},
we see that the evolutions in the two frames are related by
\begin{equation}
\partial_{t}g^{\prime}\left(\boldsymbol{r}^{\prime},\Gamma^{\prime}
;t\right)\Big|_{\boldsymbol{r}^{\prime}
=\boldsymbol{r}^{\prime}\left[\boldsymbol{r},t\right]}=\partial_{t}
g\left(\boldsymbol{r},\Gamma;t\right)+\boldsymbol{v}^{\mathrm{solv}}
\left(\boldsymbol{r}_{o}(t),t\right)\cdot\partial_{\boldsymbol{r}}
g\left(\boldsymbol{r},\Gamma;t\right).\label{eq:dt_gs}
\end{equation}

What is the advantage of switching to the coordinates in the auxiliary, then?
First, for any (potentially time-dependent) evaluation
point $\boldsymbol{r}^{\prime}$, the new Smoluchowski operator $\Omega^{\dagger\,\prime}(\Gamma^{\prime},t)$
is insensitive to global, potentially
time-dependent translations in the original frame, \emph{i.e.}, offsets
of the velocity field $\boldsymbol{v}^{\mathrm{solv}}\left(\boldsymbol{r_{i}},t_{1}\right)$.
Accordingly, it only depends on the velocity gradient 
$\kappa_{\alpha\beta}\left(\boldsymbol{r},t\right)\equiv\partial_{\beta}v^{
\mathrm{solv}} _ {
\alpha}\left(\boldsymbol{r},t\right)=\partial_{\beta}^{\prime}v_{\alpha}
^{\prime}\left(\boldsymbol{r}^{\prime}[\boldsymbol{r},t],t\right)$,
responsible for the deformation of the structure. A second considerable
benefit emerges for the specific point of evaluation $\boldsymbol{r}^{\prime}=\boldsymbol{r}_{o}$, where the
field $\boldsymbol{v}^{\prime}$ vanishes at all times. (Note that in the
\emph{original} frame
this point moves as time passes). Then, in Eq.~\ref{eq:dt_gs},
the effects of advection and of the intrinsic dynamics are clearly
separated: the latter are reflected by the evolution of $g^{\prime}$
at the point $\boldsymbol{r}^{\prime}=\boldsymbol{r}_{o}$ in the auxiliary
frame,
where there is no flow, while the second term on the right-hand side (rhs) is the desired
advection term; let us once again emphasise that this non-local term
is physically crucial in a heterogeneous system.

\subsection{Leading-order locally homogeneous flow\label{sub:Leading-order-locally-homogeneou}}

Formally, Eq.~\ref{eq:g_t_formal_sol} conveys the impression that
a microscopic observable $g$, albeit evaluated at a given point $\boldsymbol{r}_{o}$
and time $t_{o}$, depends on the configuration $\Gamma$ of all particles
\emph{throughout space}, and not only at $\boldsymbol{r}=\boldsymbol{r}_{o}$,
and hence requires the knowledge of the whole solvent velocity field
$\boldsymbol{v}^{\mathrm{solv}}\left(\boldsymbol{r},t\leqslant t_{o}\right)$.
However, making use of the short range of usual observables, we purport
to bolster the intuition that, to leading order, $\left\langle g(\boldsymbol{r}_{0})\right\rangle _{t_{o}}$
is mainly determined by the history of the velocity gradient
$\boldsymbol{\kappa}(\boldsymbol{r}_{0}(t),t)$
along the solvent pathline $\boldsymbol{r}_{o}(t)$, with
$\boldsymbol{r}(t_{o})=\boldsymbol{r}_{o}$.

We define the range of a microscopic observable $g(\boldsymbol{r})$
as the distance beyond which the particle configuration becomes irrelevant.
More precisely, $\mathrm{range}(g)$ is the minimal radius of a disk
$\mathcal{D}$ centred at $\boldsymbol{r}$ such that, for any two particle
configurations $\Gamma^{(A)}$ and $\Gamma^{(B)}$ coinciding over $\mathcal{D}$, \emph{i.e.}, such that
\begin{equation}
\boldsymbol{r_{i}}^{(A)}=\boldsymbol{r_{i}}^{(B)} \text{ if } \boldsymbol{r_{i}}^{(A)}\in\mathcal{D} \text{ or } \boldsymbol{r_{i}}^{(B)}\in\mathcal{D},
\end{equation}
 $g(\boldsymbol{r},\Gamma^{(A)})$ and $g(\boldsymbol{r},\Gamma^{(B)})$
are equal, to a good approximation. For example, the range of the density observable
$\rho(\boldsymbol{r}) \equiv \sum_{j=1}^N
\delta(\boldsymbol{r}-\boldsymbol{r_j})$ is $0^{+}$ and that of the stress
$\boldsymbol{\sigma}(\boldsymbol{r})$
is bounded by the cut-off distance of interparticle interactions. 

If the range of an observable $g(\boldsymbol{r}_{o})$ is small
compared to the lengthscale $l\sim\nicefrac{\boldsymbol{\kappa}}{\nabla\boldsymbol{\kappa}}$
over which the velocity gradient varies, we are tempted to replace
the global inhomogeneous flow with a much more tractable affine (\emph{i.e.},
homogeneous) velocity
field that coincides with the inhomogeneous one around $\boldsymbol{r}_{o}$.

This comes down to approximating the genuine Smoluchowski operator
(appearing, \emph{e.g.}, in Eq.~\ref{eq:g_t_formal_sol}) with
\begin{eqnarray*}
\Omega_{\mathrm{hom}}^{\dagger}(\Gamma;t) & = &
\Omega_{\mathrm{eq}}^{\dagger}(\Gamma)+\sum_{i=1}^{N}\left[\boldsymbol{v}^{
\mathrm{solv}}\left(\boldsymbol{r}_{o},t\right)+\boldsymbol{\kappa}
\left(\boldsymbol{r}_{o},t\right)\cdot\left(\boldsymbol{r_{i}}-\boldsymbol{r}_{o
}\right)\right]\cdot\partial_{i},
\end{eqnarray*}
where
\begin{equation}\label{omegaeq}
\Omega_{\mathrm{eq}}^{\dagger}(\Gamma)
\equiv
\sum_{i=1}^N \left[ \partial_{i}+\boldsymbol{F_{i}}\left(\Gamma\right)  \right] \cdot \partial_{i}
.
\end{equation}
How large is the error due to this approximation? At time $t$, the error reads
\begin{eqnarray*}
\left(e_{-}^{\int_{0}^{t}\Omega^{\dagger}(s)ds}-e_{-}^{\int_{0}^{t}\Omega_{
\mathrm{hom}}^{\dagger}(s)ds}\right)g(\boldsymbol{r}_{o},\Gamma).
 \end{eqnarray*}
In particular, at $t=0$, it is zero, and the first-order
term in $t$ yields its initial growth rate, 
\begin{eqnarray*}
\frac{1}{t}\int_{0}^{t}\left[\Omega^{\dagger}(s)-\Omega_{\mathrm{hom}}^{\dagger}
(s)\right]g(\boldsymbol{r}_{o},\Gamma)ds
& = &
\mathcal{O}\left(\sum_{i=1}^{N}\left\Vert \nabla\boldsymbol{\kappa}\right\Vert
\left\Vert \boldsymbol{r_{i}}-\boldsymbol{r}_{o}\right\Vert
^{2}|\partial_{i}g(\boldsymbol{r}_{o},\Gamma)|\right)\\
 & = & \mathcal{O}\left(N_{\mathcal{D}}(\Gamma)\left\Vert
\nabla\boldsymbol{\kappa}\right\Vert
|\mathrm{range}(g)|^{2}\,\underset{i\in\{1,...,N\}}{\mathrm{max}}\left\Vert
\partial_{i}g(\boldsymbol{r}_{o},\Gamma))\right\Vert \right),
\end{eqnarray*}
where $N_{\mathcal{D}}(\Gamma)$ is the number of particles within
disk $\mathcal{D}$ in configuration $\Gamma$. Clearly, $\left\Vert
\nabla\boldsymbol{\kappa}\right\Vert |\mathrm{range}(g)|^{2}$
arises because of the local deviations from affinity.
The second-order term (quadratic in $\Omega^{\dagger}$) in the
expansion of the approximation error also contains contributions
in $\left\Vert \nabla\boldsymbol{\kappa}\right\Vert |\mathrm{range}(g)|^{2}$;
some are multiplied by $\Omega^\dagger_{\mathrm{eq}}$ (which tends to restore
the equilibrium configuration), while the others involve
$\boldsymbol{v}^{\mathrm{solv}}\left(\boldsymbol{r}_{o},s\right)$, which drags
particles away. Indeed,
through advection with the solvent velocity, some particles, which initially
lay far from $\boldsymbol{r}_o$ (where the affine approximation is poor),
will enter the region $\mathcal{D}$ where they become relevant for the
computation of $g$.
Consequently, for the approximation to work best, particles close to
$\boldsymbol{r}_o$ should move as little as possible. This is exactly why 
it is advantageous to switch to the auxiliary frame introduced in the
previous section: the auxiliary driving field
$\boldsymbol{v}^{\prime}\left(\boldsymbol{r}^{\prime},t\right)$
vanishes at point $\boldsymbol{r}^{\prime}=\boldsymbol{r}_o$ at all
times\footnote{Note, however,
that the quality of this locally homogeneous approximation will dramatically
worsen
with the duration of the memory of the system and the magnitude of
$\boldsymbol{\kappa}$.}. In
that frame, the approximate evolution is ruled by

\begin{equation}
g^\prime(\boldsymbol{r}_o,\Gamma^\prime;t_0)
=
e_{-}^{\int_{0}^{t_0}\Omega_\mathrm{hom}^ {\dagger\,\prime}
(\Gamma^\prime;s)ds}
g(\boldsymbol{r}_o,\Gamma^\prime;0), 
\label{eq:loc_hom_flow}
\end{equation}
with
\[
\Omega_\mathrm{hom}^{\dagger\,\prime}(\Gamma^{\prime},t)
\equiv
\sum_{i=1}^{N}
\left[
\partial_{i}^{\prime}+\boldsymbol{F_{i}}^{\prime}\left(\Gamma^{\prime}
\right)+\boldsymbol{ \kappa}^{\prime}\left(\boldsymbol{r}_0,t\right)
\cdot(\boldsymbol
{r_i}^{\prime}-\boldsymbol {r}_0)
\right]
\cdot\partial_{i}^{\prime}.
\]
Recalling that, by definition, the original frame and the auxiliary one
coincide at time $t_0$ and thus  $\left\langle
g^\prime(\boldsymbol{r}_o)\right\rangle _{t_{o}}=\left\langle
g(\boldsymbol{r}_{o})\right\rangle _{t_{o}}$, we can now come back to the
original frame, using the change of coordinates of Eq.~\ref{eq:change_of_coord},
and confirm the intuition that, to leading order, $\left\langle
g(\boldsymbol{r}_o)\right\rangle _{t_o}$
is governed by the velocity gradient along the pathline, \emph{i.e.}, $\left\{
\boldsymbol{\kappa}\left(\boldsymbol{r}_{o}(t),t\right)\right\} $. This is the
intrinsic part of the dynamics. As a reminder, the time derivative of
$\left\langle g(\boldsymbol{r}_{o})\right\rangle _{t_{o}}$ also involves an
extrinsic part, namely, the
advection term in Eq.~\ref{eq:dt_gs}.

To proceed, physically motivated approximations, expressed as projections
onto relevant variables, are performed onto the intrinsic dynamics. In the
end, these
approximations, conducted in Fourier space, shall heavily rely on the
possibility to
treat the driving flow as (almost) \emph{globally} homogeneous, whereas the
system under study may be globally very heterogeneous.   The problem is
solved by performing these approximations in the homogeneous
auxiliary system of Eq.~\ref{eq:loc_hom_flow}, which is a reasonable surrogate
for the original one if the flow is \emph{locally} homogeneous. This is a first
step towards a systematic expansion of the velocity field in the auxiliary
frame, starting (in this paper) with a uniform velocity gradient
$\boldsymbol{\kappa}(\boldsymbol{r},t)=
\boldsymbol{\kappa}\left(\boldsymbol{r}_0(t), t\right)$, then considering the
gradient $\boldsymbol{\nabla\kappa}$ of $\boldsymbol{\kappa}$ at
$\boldsymbol{r}_0(t)$, etc.

\section{Projection Scheme}

\subsection{Sets of slow variables\label{sub:Choice-of-slow}}

In a typical mode-coupling spirit, the slow intrinsic evolution (with respect
to microscopic time scales) of a generic observable $g$ such as
the density or the stress will be captured \emph{via} its projection
onto (\emph{i.e.}, cross-correlation with) familiar slow modes. The observables
will be expressed in Fourier space,
where the collective dynamics are best captured. Since the global density
$\rho_{\boldsymbol{q}=0}$
(where $\boldsymbol{q}$ represents a wavevector in Fourier space)
is the only conserved quantity in the problem, \emph{i.e.}, $\partial_{t}\rho_{\boldsymbol{q}=0}=0$,
and the relaxation time of $\rho_{\boldsymbol{q}}$ diverges in the
limit of small $\boldsymbol{q}$, we define the linear density modes
$\left\{ \rho_{\boldsymbol{q}},\,\boldsymbol{q}\in\mathbb{R}^{d}\right\} $
in Fourier space as a first set of slow modes, associated to the projector
$P_{1}$, \emph{viz.},
\[
P_{1}\equiv\sum_{\boldsymbol{q}}\rho_{\boldsymbol{q}}\rangle\frac{1}{NS_{\boldsymbol{q}}}\langle\rho_{\boldsymbol{q}}^{\star},
\]
where $S_{\boldsymbol{q}}\equiv N^{-1}\left\langle \rho_{\boldsymbol{q}}^{\star}\rho_{\boldsymbol{q}}\right\rangle $
is the static structure factor, and its complementary part $Q_{1}\equiv1-P_{1}$.
It should be noted that the ensemble average in the projection is
performed with respect to the equilibrium distribution $\psi_{\mathrm{eq}}$
(denoted by $\left\langle \cdot\right\rangle $ here), whereas
averages over the initial distribution $\psi_{0}$ shall be denoted by
$\left\langle \cdot\right\rangle _{0}$,
\[
P_{1}g\left(\Gamma\right)=\frac{\rho_{\boldsymbol{q}}\left(\Gamma\right)}{NS_{q}}\left[\int\rho_{\boldsymbol{q}}^{\star}\left(\Gamma^{\prime}\right)g\left(\Gamma^{\prime}\right)\psi_{\mathrm{eq}}\left(\Gamma^{\prime}\right)d\Gamma^{\prime}\right].
\]
In Ref.~\cite{Brader2012}, Brader \emph{et
al.} noticed the absence of any coupling with
linear density modes in a purely homogeneous flow and thus further projected
the dynamics onto density
pairs $\left\{
\rho_{\boldsymbol{k}}\rho_{\boldsymbol{q}},\,\boldsymbol{k},\,\boldsymbol{q}
\in\mathbb{R}^{d}\right\} $ with the projector 
\[
P_{2}\equiv\sum_{\boldsymbol{k}>\boldsymbol{q}}\rho_{\boldsymbol{k}}\rho_{\boldsymbol{q}}\rangle\frac{1}{N^{2}S_{\boldsymbol{k}}S_{\boldsymbol{q}}}\langle\rho_{\boldsymbol{k}}^{\star}\rho_{\boldsymbol{q}}^{\star},
\]
where the Gaussian approximation $\left\langle
\rho_{\boldsymbol{k}}^{\star}\rho_{\boldsymbol{q}}^{\star}\rho_{\boldsymbol{k}}
\rho_{\boldsymbol{q}}\right\rangle \approx\left\langle
\rho_{\boldsymbol{k}}^{\star}\rho_{\boldsymbol{k}}\right\rangle \left\langle
\rho_{\boldsymbol{q}}^{\star}\rho_{\boldsymbol{q}}\right\rangle
=N^{2}S_{\boldsymbol{k}}S_{\boldsymbol{q}}$ was used, and its complementary part
$Q_{2}$. Although
this section comes in the wake of Ref.~\cite{Brader2012}, we shall
not neglect the couplings with linear density modes from the outset,
because the flow is not strictly homogeneous.

\subsection{Generalised Green-Kubo relation}

To prepare the projection, we recast Eq.~\ref{eq:g_t_formal_sol} into a form
which better highlights the deviations from the initial configuration occurring
throughout the past.

In the Schr\"odinger-like formulation, we denote by $\delta\psi$ these
flow-induced perturbations, \emph{viz}., 
\[
\psi(\Gamma,t)=\psi_{0}(\Gamma)+\delta\psi(\Gamma,t).
\]
Since $\partial_{t}\psi(\Gamma,t)=\Omega(\Gamma,t)\left[\psi_{0}(\Gamma)+\delta\psi(\Gamma,t)\right]$,
solving for $\delta\psi$ yields
\begin{equation}
\psi(\Gamma,t)=\psi_{0}(\Gamma)+\int_{0}^{t}dt_{1}e_{+}^{\int_{t_{1}}^{t}
\Omega(\Gamma,s)ds}\Omega(\Gamma
,t_{1})\psi_{0}(\Gamma).
\label{eq:psi_omega_1}
\end{equation}
  For the time being, $\Omega$ is the Smoluchowski operator for a generic
flow field $\boldsymbol{v}^{\mathrm{solv}}$, but the approximations performed in
the following (see Section \ref{sec:red_loc_hom}) will hinge on its being close
to homogeneous, so one might already think of $\Omega$  and $\Omega^\dagger$ as
their homogeneous auxiliary-frame surrogates of Eq.~\ref{eq:loc_hom_flow},
that is to say, mentally consider the evolution in the auxiliary
frame, with the replacement $\Omega^\dagger \rightarrow
\Omega_\mathrm{hom}^{\dagger\,\prime}$.

Applying Eq.~\ref{eq:psi_omega_1}  to an arbitrary observable $g$,
\emph{e.g.},
$g=\boldsymbol{\sigma}$, and partially integrating the time-ordered
exponential, we arrive at a generalised Green-Kubo (gGK) relation, expressed in
the Heisenberg-like representation,
\begin{eqnarray}
\left\langle g\right\rangle _{t} & = & \int d\Gamma g\left(\Gamma\right)\left[\psi_{0}(\Gamma)+\int d\Gamma\int_{0}^{t}dt_{1}e_{+}^{\int_{t_{1}}^{t}\Omega(\Gamma,s)ds}\Omega(\Gamma,t_{1})\psi_{0}(\Gamma)\right]\nonumber \\
 & = & \left\langle g\right\rangle _{0}+\int_{0}^{t}dt_{1}\left\langle
\Omega^{\dagger}(t_{1})e_{-}^{\int_{t_{1}}^{t}\Omega^{\dagger}(s)ds}
g\right\rangle _{0}.\label{eq:Gen_Green_Kubo}
\end{eqnarray}

The Green-Kubo nature of Eq.~\ref{eq:Gen_Green_Kubo} becomes clearer
if the integrand is rewritten as 
\[
\int
d\Gamma \left[\Omega(\Gamma,t_{1})\psi_{0}(\Gamma)\right]e_{-}^{\int_{t_{1}}^{t}
\Omega^{\dagger}(s)ds}g.
\]
$\Omega(\Gamma,t_{1})\psi_{0}(\Gamma)$ is thus the deviation from
$\psi_{0}(\Gamma)$ created at time $t_{1}$ (per unit time). For
instance, for simple shear flow, starting with $\psi_{0}=\psi_{\mathrm{eq}}$, the deviation couples strain rate and stress:
$\Omega(\Gamma,t_{1})\psi_{\mathrm{eq}}(\Gamma)=\dot{\gamma}(t_1)\sigma_{xy}
(\Gamma)\psi_{\mathrm{eq}}(\Gamma)$,
where  $\sigma_{xy}$ is the shear element of the Kirkwood  stress tensor and $\dot{\gamma}(t)$ is the imposed shear rate.

\subsection{Projected dynamics}

\begin{figure}
\begin{centering}
\includegraphics[width=5cm]{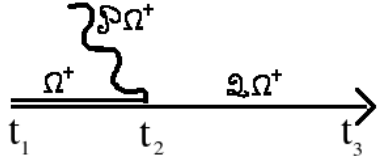}
\par\end{centering}

\caption{\label{fig:Schematic-diagram-of}Schematic diagram of the decomposition
of the full propagator $U^{\dagger}(t,t_{1})$, associated with the
operator $\Omega^{\dagger}$, as performed in Eq.~\ref{eq:full_prop_dec}.}
\end{figure}

Let $U^{\dagger}(t,t_{1})=e^{\int_{t_{1}}^{t}dt_{2}\Omega^{\dagger}(t_{2})}$
be the propagator appearing in gGK (Eq.~\ref{eq:Gen_Green_Kubo}),
associated with the full dynamics $\Omega^{\dagger}$. We split $U^{\dagger}$
into a part $U_{1}^{\dagger}(t,t_{1})\equiv e^{\int_{t_{1}}^{t}dt_{1}Q_{1}\Omega^{\dagger}(t_{1})}$
that evolves purely orthogonally to $P_{1}$ and a part that interacts
at least once with $P_{1}$ (the notation $t_{2}$ referring to the
time of the last interaction, see Fig.~\ref{fig:Schematic-diagram-of}),
\emph{viz.}
\begin{equation}
U^{\dagger}(t,t_{1})=\int_{t_{1}}^{t}dt_{2}U^{\dagger}(t_{2},t_{1})P_{1}\Omega^{\dagger}(t_{2})U_{1}^{\dagger}(t,t_{2})+U_{1}^{\dagger}(t,t_{1}).\label{eq:full_prop_dec}
\end{equation}

Inserting the decomposition of Eq.~\ref{eq:full_prop_dec}
into gGK (Eq.~\ref{eq:Gen_Green_Kubo}), we arrive at:
\begin{eqnarray}
\left\langle g\right\rangle _{t}-\left\langle g\right\rangle _{0} & = & \int_{0}^{t}dt_{1}\left\langle \Omega^{\dagger}(t_{1})U^{\dagger}(t,t_{1})g\right\rangle _{0}\nonumber \\
 & = & \int_{0}^{t}dt_{2}\int_{0}^{t_{1}}dt_{1}\left\langle \Omega^{\dagger}(t_{1})U^{\dagger}(t_{2},t_{1})P_{1}\Omega^{\dagger}(t_{2})U_{1}^{\dagger}(t,t_{2})g\right\rangle _{0}\nonumber \\
 &  & +\int_{0}^{t}dt_{1}\left\langle \Omega^{\dagger}(t_{1})U_{1}^{\dagger}(t,t_{1})g\right\rangle _{0}\nonumber \\
 & = & \notag \int_{0}^{t}dt_{2}\sum_{\boldsymbol{q}}\underset{\left\langle
\rho_{\boldsymbol{q}}\right\rangle _{t_{2}}-\left\langle
\rho_{\boldsymbol{q}}\right\rangle
_{0}}{\underbrace{\int_{0}^{t_{2}}dt_{1}\left\langle
\Omega^{\dagger}(t_{1})U^{\dagger}(t_{2},t_{1})\rho_{\boldsymbol{q}}
\right\rangle _{0}}}\frac{\left\langle
\rho_{\boldsymbol{q}}^{\star}\Omega^{\dagger}(t_{2})U_{1}^{\dagger}(t,t_{2}
)g\right\rangle }{NS_{\boldsymbol{q}}}\label{eq:g_t_Orth1}\\
 &  & +\underset{\mathrm{(Orth_{1})}}{\underbrace{\int_{0}^{t}dt_{1}\left\langle \Omega^{\dagger}(t_{1})U_{1}^{\dagger}(t,t_{1})g\right\rangle _{0}}}
\end{eqnarray}
where we have made use of gGK (applied to density fluctuations, $g \to \rho_{\boldsymbol{q}}$) in the last equality to reduce the first
brace to $\left\langle \rho_{\boldsymbol{q}}\right\rangle _{t_{2}}-\left\langle \rho_{\boldsymbol{q}}\right\rangle _{0}$.
One thus arrives at:
\begin{equation}
\left\langle g\right\rangle _{t}-\left\langle g\right\rangle _{0} =
\int_{0}^{t}dt_{2}\sum_{\boldsymbol{q}}\left(\left\langle
\rho_{\boldsymbol{q}}\right\rangle _{t_{2}}-\left\langle
\rho_{\boldsymbol{q}}\right\rangle _{0}\right)\frac{\left\langle
\rho_{\boldsymbol{q}}^{\star}\Omega^{\dagger}(t_{2})U_{1}^{\dagger}(t,t_{2}
)g\right\rangle }{NS_{\boldsymbol{q}}}+\mathrm{(Orth_{1})}.
\label{eq:g_t_general}
\end{equation}

First, we focus on the dynamical correlator $\left\langle \rho_{\boldsymbol{q}}^{\star}\Omega^{\dagger}(t_{2})U_{1}^{\dagger}(t,t_{2})g\right\rangle $
on the rhs and introduce the identity $P_{1}+Q_{1}=1$
as follows:
\begin{eqnarray*}
\frac{\left\langle \rho_{\boldsymbol{q}}^{\star}\Omega^{\dagger}(t_{2})U_{1}^{\dagger}(t,t_{2})g\right\rangle }{NS_{\boldsymbol{q}}} & = & \frac{1}{NS_{\boldsymbol{q}}}\left\langle \rho_{\boldsymbol{q}}^{\star}\Omega^{\dagger}(t_{2})U_{1}^{\dagger}(t,t_{2})\left(P_{1}+Q_{1}\right)g\right\rangle \\
 & = & -\sum_{\boldsymbol{k}}V_{\boldsymbol{k}}^{g}M_{\boldsymbol{qk}}^{(1)}\left(t,t_{2}\right)\\
 &  & +\frac{1}{NS_{\boldsymbol{q}}}\left\langle \rho_{\boldsymbol{q}}^{\star}\Omega^{\dagger}(t_{2})U_{1}^{\dagger}(t,t_{2})Q_{1}g\right\rangle ,
\end{eqnarray*}
where the vertex $V_{\boldsymbol{k}}^{g}\equiv\frac{\left\langle \rho_{\boldsymbol{k}}^{\star}g\right\rangle }{NS_{\boldsymbol{k}}}$
quantifies the coupling of the observable $g$ to the density mode
$\rho_{\boldsymbol{k}}$ in the equilibrium distribution and 
\begin{eqnarray}\label{eq:m}
M_{\boldsymbol{qk}}^{(1)}\left(t,t_{2}\right) & \equiv & -\frac{\left\langle \rho_{\boldsymbol{q}}^{\star}\Omega^{\dagger}(t_{2})U_{1}^{\dagger}(t,t_{2}) \rho_{\boldsymbol{k}}\right\rangle }{NS_{\boldsymbol{q}}}\\
 & = &\frac{1}{NS_{\boldsymbol{q}}}
\Big\langle
\Big[
\rho_{\boldsymbol{q}}^{\star} \sum_{j}  \boldsymbol{v}^{\mathrm{solv}}
(\boldsymbol{r}_{j},t_{2}) \cdot 
\boldsymbol{F}_{j}
 \notag
\\
 &  & \,\,\,\,+
 i\boldsymbol{q}\cdot
\left( - \boldsymbol{\hat{F}_q}^\star + \boldsymbol{\hat{v}_q}^\star(t_2)
\right)
\Big]
U_{1}^{\dagger}(t,t_{2})
\rho_{\boldsymbol{k}}
\Big\rangle \notag
\end{eqnarray}
is a memory kernel evaluated in the equilibrium distribution, with

\begin{equation}
\label{eq:F_and_v_hats}
 \boldsymbol{\hat{F}_q} \equiv \sum_{j}
\boldsymbol{F}_{j} e^{-i\boldsymbol{q}\cdot\boldsymbol{r}_{j}}
\text{ and }
 \boldsymbol{\hat{v}_q}(t) \equiv \sum_{j}
\boldsymbol{v}^{\mathrm{
solv}}(\boldsymbol{r}_{j},t) e^{-i\boldsymbol{q}\cdot\boldsymbol{r}_{j}}.
\end{equation}

\subsection{Application to the density observable}

Before turning to our main interest, \emph{i.e.}, the stress, we wish
to illustrate the principle of the projection scheme for a generic flow, but on
a simpler
observable, namely, the density $g=\rho_{\boldsymbol{p}},\,\boldsymbol{p}\in\mathbb{R}^{d}$,
for which the complement $Q_{1}g$ vanishes by definition. The following
calculations need not be performed in the homogeneous auxiliary frame; they
hold true for an inhomogeneous flow.

Applying Eq.~\ref{eq:g_t_general}  
to density
modes ($V_{\boldsymbol{k}}^{\rho_{\boldsymbol{p}}}\equiv\frac{\left\langle
\rho_{\boldsymbol{k}}^{\star}\rho_{\boldsymbol{p}}\right\rangle
}{NS_{\boldsymbol{k}}}=\delta_{\boldsymbol{k},\boldsymbol{p}}$)
leads to
\begin{eqnarray*}
\left\langle \rho_{\boldsymbol{p}}\right\rangle _{t}-\left\langle
\rho_{\boldsymbol{p}}\right\rangle _{0} & = &
-\int_{0}^{t}dt_{1}\sum_{\boldsymbol{q}}\left[\left\langle
\rho_{\boldsymbol{q}}\right\rangle _{t_{1}}-\left\langle
\rho_{\boldsymbol{q}}\right\rangle
_{0}\right]M_{\boldsymbol{qp}}^{(1)}\left(t,t_{1}\right)+(\mathrm{Orth}1),
\end{eqnarray*}
where $M_{\boldsymbol{qp}}^{(1)}\left(t,t_{1}\right)$
is given in Eq.~\ref{eq:m}
and $(\mathrm{Orth}1)\equiv\int_{0}^{t}dt_{1}\left\langle \Omega^{\dagger}(t_{1})U_{1}^{\dagger}(t,t_{1})\rho_{\boldsymbol{p}}\right\rangle _{0}$.

Taking a derivative with respect to time $t$ yields
\begin{eqnarray}
\partial_{t}\left\langle \rho_{\boldsymbol{p}}\right\rangle _{t} & = &
-\sum_{\boldsymbol{q}}\left[\left\langle \rho_{\boldsymbol{q}}\right\rangle
_{t}-\left\langle \rho_{\boldsymbol{q}}\right\rangle
_{0}\right]M_{\boldsymbol{qp}}^{(1)}\left(t,t\right) \nonumber
\\
&  & \:\:\:\:-\int_{0}^{t}dt_{1}
\sum_{\boldsymbol{q}}\left[\left\langle \rho_{\boldsymbol{q}}\right\rangle
_{t_{1}}-\left\langle \rho_{\boldsymbol{q}}\right\rangle
_{0}\right]\partial_{t}M_{\boldsymbol{qp}}^{(1)}(t,t_{1})+\hat{S}_{\boldsymbol{p
}}
(t).\label{eq:density_example}
\end{eqnarray}
Here, we have used the explicit notation $\hat{S}_{\boldsymbol{p}}(t)$
for $\partial_{t}\mathrm{(Orth1)}$;
\begin{equation}
\hat{S}_{\boldsymbol{p}}(t) = 
-i\boldsymbol{p} \cdot \langle \boldsymbol{\hat{F}_p} + \boldsymbol{\hat{v}_p}
- i \boldsymbol{p} \rho_{\boldsymbol{p}} \rangle_0
- \int_0^t dt_1 \langle \Omega^\dagger(t_1)
U_1^\dagger(t,t_1) Q_1 i\boldsymbol{p} \cdot( \boldsymbol{\hat{F}_p} +
\boldsymbol{\hat{v}_p}) \rangle_0.
\label{eq:Sp_t}
\end{equation}

The term
$M_{\boldsymbol{qp}}^{(1)}\left(t,t\right)=-\left(NS_{\boldsymbol{q}}
\right)^{-1}\left\langle
\rho_{\boldsymbol{q}}^{\star}\Omega^{\dagger}(t)\rho_{\boldsymbol{p}}
\right\rangle $
can be simplified. Using the equilibrium (\emph{i.e.,} $\boldsymbol{v}^{\mathrm{solv}}=\boldsymbol{0}$)
Smoluchowski operator $\Omega_{\mathrm{eq}}^{\dagger}$ from Eq.~\ref{omegaeq}, we can write
\begin{eqnarray*}
\left\langle \rho_{\boldsymbol{q}}^{\star}\Omega^{\dagger}(t)\rho_{\boldsymbol{p}}\right\rangle  & = & \left\langle \rho_{\boldsymbol{q}}^{\star}\Omega_{\mathrm{eq}}^{\dagger}\rho_{\boldsymbol{p}}\right\rangle +\left\langle \rho_{\boldsymbol{q}}^{\star}\sum_{j}\boldsymbol{v}^{\mathrm{solv}}\left(\boldsymbol{r_{j}},t\right)\cdot\partial_{j}\rho_{\boldsymbol{p}}\right\rangle ,
\end{eqnarray*}
where, using partial integration,

\begin{eqnarray*}
\left\langle \rho_{\boldsymbol{q}}^{\star}\Omega_{\mathrm{eq}}^{\dagger}\rho_{\boldsymbol{p}}\right\rangle  & = & \left\langle \rho_{\boldsymbol{q}}^{\star}\sum_{j}\left(\partial_{j}+\boldsymbol{F_{j}}\right)\cdot\partial_{j}\rho_{\boldsymbol{p}}\right\rangle \\
 & = & \int d\Gamma\psi_{\mathrm{eq}}(\Gamma)\rho_{\boldsymbol{q}}^{\star}\sum_{j}\left(\partial_{j}+\boldsymbol{F_{j}}\right)\cdot\partial_{j}\rho_{\boldsymbol{p}}\\
 & = & -\int d\Gamma\sum_{j}\left\{ \partial_{j}\left[\psi_{\mathrm{eq}}(\Gamma)\rho_{\boldsymbol{q}}^{\star}\right]-\rho_{\boldsymbol{q}}^{\star}\sum_{j}\boldsymbol{F_{j}}\psi_{\mathrm{eq}}(\Gamma)\right\} \cdot\partial_{j}\rho_{\boldsymbol{p}}\\
 & = & -\boldsymbol{q}\cdot\boldsymbol{p}\left\langle
\sum_{j}e^{i(\boldsymbol{q}-\boldsymbol{p})\cdot\boldsymbol{r_{j}}}\right\rangle
\\
 & = & -\boldsymbol{p}^{2}\delta_{\boldsymbol{qp}}N.
\end{eqnarray*}
The second term, $\left\langle \rho_{\boldsymbol{q}}^{\star}\sum_{j}\boldsymbol{v}^{\mathrm{solv}}\left(\boldsymbol{r_{j}},t\right)\cdot\partial_{j}\rho_{\boldsymbol{p}}\right\rangle \equiv F_{\boldsymbol{qp}}$,
is most easily simplified by first backward-Fourier transforming
it \emph{with respect to $\boldsymbol{p}$ only},\emph{ viz.}, 
\begin{eqnarray*}
F_{\boldsymbol{q}}\left(\boldsymbol{r}_{0}\right) & = & \left\langle
\rho_{\boldsymbol{q}}^{\star}\sum_{j}\boldsymbol{v}^{\mathrm{solv}}
\left(\boldsymbol{r_{j}},t\right)\cdot\partial_{j}\rho\left(\boldsymbol{r}_{0}
\right)\right\rangle ,
\\
 & = & -\left\langle \rho_{\boldsymbol{q}}^{\star}\sum_{j}\boldsymbol{v}^{\mathrm{solv}}\left(\boldsymbol{r_{j}},t\right)\cdot\partial_{\boldsymbol{r}_{0}}\delta\left(\boldsymbol{r}_{0}-\boldsymbol{r_{j}}\right)\right\rangle .
\\
 & = & -\partial_{\boldsymbol{r}_{0}} \cdot \left\langle
\rho_{\boldsymbol{q}}^{\star}\sum_{j}\boldsymbol{v}^{\mathrm{solv}}
\left(\boldsymbol{r_{j}},t\right)
\delta\left(\boldsymbol{r}_{0}-\boldsymbol{r_{j}}\right)\right\rangle
\\
 & = & -\left\langle
\rho_{\boldsymbol{q}}^{\star}\boldsymbol{v}^{\mathrm{solv}}\left(\boldsymbol{r}_
{0},t\right)\cdot\partial_{\boldsymbol{r}_{0}}\rho\left(\boldsymbol{r}_{0}
\right)\right\rangle,
\end{eqnarray*}
where we have used the incompressibility of the velocity field, \emph{viz}.,
$\partial_{\boldsymbol{r}_{0}}\cdot\boldsymbol{v}^{\mathrm{solv}}\left(\boldsymbol{r}_{0},t\right)=0$.
It suffices to transform $F_{\boldsymbol{q}}\left(\boldsymbol{r}_{0}\right)$
back into reciprocal space (with respect to $\boldsymbol{r}_{0}$)
to obtain:
\begin{eqnarray*}
F_{\boldsymbol{qp}} & = & -\sum_{\boldsymbol{k}}\left\langle \rho_{\boldsymbol{q}}^{\star}\boldsymbol{v}_{\boldsymbol{k}}(t)\cdot i\left(\boldsymbol{p}-\boldsymbol{k}\right)\rho_{\boldsymbol{p}-\boldsymbol{k}}\right\rangle \\
 & = & -\boldsymbol{v}_{\boldsymbol{k}}(t)\cdot i\boldsymbol{q}\left\langle \rho_{\boldsymbol{q}}^{\star}\rho_{\boldsymbol{q}}\right\rangle \\
 & = & -iN\boldsymbol{v}_{\boldsymbol{p}-\boldsymbol{q}}^{\mathrm{solv}}\left(t\right)\cdot\boldsymbol{q}S_{\boldsymbol{q}}.
\end{eqnarray*}

Collecting these contributions into Eq.~\ref{eq:density_example},
one arrives at an equation of evolution of density fluctuations:

\begin{eqnarray}
\partial_{t}\left\langle \rho_{\boldsymbol{p}}\right\rangle
_{t}+\sum_{\boldsymbol{q}}\boldsymbol{v}_{\boldsymbol{p}-\boldsymbol{q}}^{
\mathrm{solv}}\left(t\right)\cdot i\boldsymbol{q}\left\langle
\rho_{\boldsymbol{q}}\right\rangle _{t}&=&
-\frac{\boldsymbol{p}^{2}}{S_{\boldsymbol{p}}}\left\langle
\rho_{\boldsymbol{p}}\right\rangle
_{t}
-\sum_{\boldsymbol{q}}\int_{0}^{t}dt_{1}\partial_{t}M_{\boldsymbol{qp}}^
{(1) }\left(t,t_{1}\right)\left\langle \rho_{\boldsymbol{q}}\right\rangle
_{t_{1}} \nonumber
\\
&  &
\:\:\:\:\:\:\:\:\:\:\:\:\:\:\:+\hat{S}_{\boldsymbol{p}}(t).
\label{eq:density_equation}
\end{eqnarray}

Let us emphasise the physical meaning of the different terms:

(i) the second term on the lhs is the advection term established in
Section~\ref{sub:Advection_term_gal};

(ii) the first one on the rhs permits the relaxation of fluctuations
through diffusion. Note that the normalising factor $S_{\boldsymbol{p}}^{-1}$
is expected, because the relaxation of a density mode $\rho_{\boldsymbol{p}}$
does not require single-particle diffusion over a lengthscale $\left\Vert
\boldsymbol{p}\right\Vert^{-1} $! Besides, if the density field is smoothed
\emph{via} coarse-graining (\emph{i.e.},
$\rho_{\boldsymbol{p}}\rightarrow\rho_{\boldsymbol{p}}\phi_{\boldsymbol{p}}$,
with $\phi_{\boldsymbol{p}}$ akin to a Gaussian of half-width a few
particle diameters), the fast relaxational modes at high
$\boldsymbol{p}$ are suppressed, and the normalising factor then tends to
$S_{\boldsymbol{0}}^{-1}$,
which is directly related to the compressibility of the suspension. 

(iii) The second term on the rhs reflects the evolution in orthogonal
space of the density fluctuations created in the past and their final
coupling back to present fluctuations. 

(iv) \emph{$\hat{S}_{\boldsymbol{p}}(t)$ }is a source term that
results from interactions with nonlinear density modes. Consistently
with the expectation that density heterogeneities in an incompressible
flow are due to collective effects, \emph{e.g.}, stress equilibration,
it is the only term that can potentially create density inhomogeneities,
insofar as the other terms are associated to pre-existing density
heterogeneities. If a homogeneous flow is imposed to an initally uniform
system, translational invariance imposes that, for finite $\boldsymbol{p}$,
$\left\langle \boldsymbol{\hat{F}_p}
\right\rangle_0 = 0$, $\left\langle \boldsymbol{\rho_p} \right\rangle_0 = 0$,
$i\boldsymbol{p} \cdot \left\langle \boldsymbol{\hat{v}_p}
\right\rangle_0 = 0$
in Eq.~\ref{eq:Sp_t}, \emph{ergo} $\hat{S}_{\boldsymbol{p}}(t)=0$ at all times,
for all $\boldsymbol{p}$; as
expected,
the source term then vanishes.

Finally, by comparing  Eq.~\ref{eq:density_equation} to the mass conservation
equation,
\[
\partial_t \langle \rho_{\boldsymbol{p}} \rangle_t + i\boldsymbol{p}\cdot
\langle \boldsymbol{j}^{\mathrm{coll}}_{\boldsymbol{p}} \rangle_t = 0,
\]
where $ \boldsymbol{j}^{\mathrm{coll}}_{\boldsymbol{p}} \equiv
 \sum_{j=1}^N \boldsymbol{\dot{r}_j} e^{-i\boldsymbol{p}\cdot
\boldsymbol{r_j}}$ is the colloidal flux, it can be seen that, in a
heterogeneous flow, the colloidal velocity
$\boldsymbol{u}^{\mathrm{coll}}(\boldsymbol{r},t)\equiv
\frac
{\langle \boldsymbol{j}^{\mathrm{coll}}(\boldsymbol{r}) \rangle_t}
{\langle \rho(\boldsymbol{r}) \rangle_t}$ will generally differ from the
driving solvent velocity $\boldsymbol{v}^{\mathrm{solv}}(\boldsymbol{r},t)$.

\subsection{Orthogonal dynamics}

Let us come back to a generic observable $g$ and refine the description
of term (Orth1) in Eq.~\ref{eq:g_t_Orth1}. In principle, the propagator
decomposition of Eq.~\ref{eq:full_prop_dec} can be iterated, and
the propagator $U_{1}^{\dagger}$, split into a part evolving with
$P_{2}Q_{1}\Omega^{\dagger}$ and an orthogonal part $U_{2}^{\dagger}$,
and so on, \emph{ad libitum}. Schematically, one would then get
\begin{eqnarray}
\left\langle g\right\rangle _{t} & \approx & \left\langle g\right\rangle
_{0}-\int_{0}^{t}dt_{2}\sum_{\boldsymbol{q},\boldsymbol{k}}\left(\left\langle
\rho_{\boldsymbol{q}}\right\rangle _{t_{2}}-\left\langle
\rho_{\boldsymbol{q}}\right\rangle
_{0}\right)V_{1}\left(\cdot\right)M_{\boldsymbol{qk}}^{(1)}\left(t,t_{2}
\right)\label{eq:seq_proj}\\
 &  &
-\int_{0}^{t}dt_{2}\sum_{\boldsymbol{k},\boldsymbol{p},\boldsymbol{k^\prime},
\boldsymbol{p^\prime}}\left(\left\langle
\rho_{\boldsymbol{k}}\rho_{\boldsymbol{p}}\right\rangle _{t_{2}}-\left\langle
\rho_{\boldsymbol{k}}\rho_{\boldsymbol{p}}\right\rangle
_{0}\right)V_{2}\left(\cdot\right)M_{\boldsymbol{qkq^{\prime}k^{\prime}}}^{(2)}
(t,t_{2})+\ldots, \notag
\end{eqnarray}
and the orthogonal evolutions denoted by $M^{(n)}(t,t_{2})$ would
then be constrained to smaller and smaller spaces. However, following
Ref.~\cite{Brader2012}, we adopt a more pragmatic approach by directly
introducing the projector $P_{2}$ in (Orth1), \emph{viz.,}
\begin{eqnarray}
\mathrm{(Orth_{1})} & \equiv & \int_{0}^{t}dt_{1}\left\langle \Omega^{\dagger}(t_{1})Q_{1}U_{1}^{\dagger}(t,t_{1})g\right\rangle _{0}\nonumber \\
 & \approx & \int_{0}^{t}dt_{1}\left\langle \Omega^{\dagger}(t_{1})Q_{1}P_{2}U_{1}^{\dagger}(t,t_{1})P_{2}g\right\rangle _{0}\nonumber \\
 & \approx & \int_{0}^{t}dt_{1}\underset{_{\boldsymbol{k^{\prime}}>\boldsymbol{p^{\prime}}}}{\sum_{\boldsymbol{k}>\boldsymbol{p}}}V_{0,\boldsymbol{k},\boldsymbol{p}}\left(t_{1}\right)V_{\boldsymbol{k^{\prime}},\boldsymbol{p^{\prime}}}^{g}\frac{\left\langle \rho_{\boldsymbol{k}}^{\star}\rho_{\boldsymbol{p}}^{\star}U_{1}^{\dagger}(t,t_{1})\rho_{\boldsymbol{k^{\prime}}}\rho_{\boldsymbol{p^{\prime}}}\right\rangle }{N^{2}S_{\boldsymbol{k}}S_{\boldsymbol{p}}},\label{eq:Orth1}
\end{eqnarray}
where 
\begin{eqnarray*}
V_{0,\boldsymbol{k},\boldsymbol{p}}\left(t_{1}\right) & \equiv & \left\langle \Omega^{\dagger}(t_{1})Q_{1}\rho_{\boldsymbol{k}}\rho_{\boldsymbol{p}}\right\rangle _{0}\\
V_{\boldsymbol{k^{\prime}},\boldsymbol{p^{\prime}}}^{g} & \equiv & \frac{\left\langle \rho^*_{\boldsymbol{k^{\prime}}}\rho^*_{\boldsymbol{p^{\prime}}}g\right\rangle }{N^{2}S_{\boldsymbol{k^{\prime}}}S_{\boldsymbol{p^{\prime}}}}
\end{eqnarray*}
are vertices that represent, respectively, the creation and relaxation of
bilinear
density modes with respect to the initial configuration and the coupling
strength of the observable $g$ to these modes.   If the
initial configuration $\psi_0$ and the velocity gradient at $t_1$ are close to
homogeneous, then $V_{0,\boldsymbol{k},\boldsymbol{p}}\left(t_{1}\right)$ is
nonzero only for $\boldsymbol{p} \approx -\boldsymbol{k} $, and if $g$ is a
spatial average, that is, if only small-wavenumber modes contribute to its
Fourier decomposition, then
$V_{\boldsymbol{k^{\prime}},\boldsymbol{p^{\prime}}}^{g}$ is nonzero only if
$\boldsymbol{p^\prime} \approx -\boldsymbol{k^\prime} $.

To conclude, the density-pair correlation function $\left\langle \rho_{\boldsymbol{k}}^{\star}\rho_{\boldsymbol{p}}^{\star}U_{1}^{\dagger}(t,t_{1})\rho_{\boldsymbol{k^{\prime}}}\rho_{\boldsymbol{p^{\prime}}}\right\rangle $
is approximated through Gaussian factoring, which is a central approximation
of mode-coupling theory \cite{Wu2003}:
\begin{eqnarray*}
\frac{\left\langle \rho_{\boldsymbol{k}}^{\star}
\rho_{\boldsymbol{p}}^{\star}U_{1}^{\dagger}(t,t_{1})\rho_{\boldsymbol{k^{\prime
}}}\rho_{\boldsymbol{p^{\prime}}}\right\rangle
}{N^{2}S_{\boldsymbol{k}}S_{\boldsymbol{p}}} & \approx &  \frac{\left\langle
\rho_{\boldsymbol{k}}^{\star}
\rho_{\boldsymbol{p}}^{\star}U^{\dagger}(t,t_{1})\rho_{\boldsymbol{k^{\prime
}}}\rho_{\boldsymbol{p^{\prime}}}\right\rangle
}{N^{2}S_{\boldsymbol{k}}S_{\boldsymbol{p}}}
\\
& \approx & \frac{\left\langle
\rho_{\boldsymbol{k}}^{\star}U^{\dagger}(t,t_{1})\rho_{\boldsymbol{k^{\prime}}}
\right\rangle }{NS_{\boldsymbol{k}}}\frac{\left\langle
\rho_{\boldsymbol{p}}^{\star}U^{\dagger}(t,t_{1})\rho_{\boldsymbol{p^{\prime}}}
\right\rangle }{NS_{\boldsymbol{p}}}\\
 & = & \Phi_{\boldsymbol{kk^{\prime}}}(t,t_{1})\Phi_{\boldsymbol{pp^{\prime}}}(t,t_{1}),
\end{eqnarray*}
where we have introduced the transient density correlator
\[
\Phi_{\boldsymbol{qk}}(t,t_{2})\equiv\frac{\left\langle \rho_{\boldsymbol{q}}^{\star}U^{\dagger}(t,t_{2})\rho_{\boldsymbol{k}}\right\rangle }{NS_{\boldsymbol{q}}}.
\]
This correlator indicates how fast density fluctuations relax, in
the presence of flow. A central result of the rheological extension
of MCT by  Brader and colleagues \cite{Brader2012}
ascertains that, for a translationally invariant (\emph{i.e.}, homogeneous)
flow, $\Phi_{\boldsymbol{qk}}(t,t_{2})$ is non-zero only if $\boldsymbol{q}$
coincides with $\boldsymbol{k}$ in the solvent flow frame\emph{,
i.e.}, if the flow advects wavevector $\boldsymbol{k}$ at time $t_{2}$
into $\boldsymbol{q}=\boldsymbol{k}(t,t_{2})\equiv e_{+}^{\int_{t_{2}}^{t}ds\boldsymbol{\kappa}(s)}$
at a later time $t$, where $\boldsymbol{\kappa}$
 is the uniform
velocity gradient (recall
$\kappa_{\alpha\beta}\left(\boldsymbol{r},t\right)\equiv\partial_{
\beta } v^{\mathrm{solv}}_ {
\alpha}\left(\boldsymbol{r},t\right)$). Interestingly, the norm of
$\boldsymbol{k}(t,t_{2})$
increases with the deformation; consequently, $\Phi$ is effectively
evaluated at a \emph{smaller} lengthscale, where thermal relaxation
occurs faster. But flow heterogeneities induce additional cross-couplings
between $\rho_{\boldsymbol{q}}^{\star}$ and $U^{\dagger}(t,t_{2})\rho_{\boldsymbol{k}}$,
through the interaction between the structure (the density modes
$\rho_{\boldsymbol{q}}$
and $\rho_{\boldsymbol{k}}$) and the flow field (the velocity gradient
modes $\boldsymbol{\kappa_{p}}(s)$).\footnote{This can be seen by
replacing $U^{\dagger}(t,t_{2})\rho_{\boldsymbol{k}}$ in
$\Phi_{\boldsymbol{qk}}(t,t_{2})$ with its gGK
expression (Eq.~\ref{eq:Gen_Green_Kubo}). The resulting integral involves a
rate of deviation $\Omega(s)\psi_{\mathrm{eq}}
=- \sum_j \boldsymbol{v}^{\mathrm{solv}}(\boldsymbol{r_j},s)\cdot
\boldsymbol{F_j} \psi_{\mathrm{eq}} $, which is proportional to
$\boldsymbol{\kappa_0}(s):\boldsymbol{\sigma_0}$ if the flow is homogeneous, but
comprises non-zero Fourier modes otherwise.}

\medskip{}

After collecting all terms, we arrive at the following expression
for Eq.~\ref{eq:g_t_Orth1}:
\begin{eqnarray}
\left\langle g\right\rangle _{t}-\left\langle g\right\rangle _{0} & \approx &
-\int_{0}^{t}dt_{1}\sum_{\boldsymbol{q}}\left(\left\langle
\rho_{\boldsymbol{q}}\right\rangle _{t_{1}}-\left\langle
\rho_{\boldsymbol{q}}\right\rangle _{0}\right) \nonumber
\\
& & \:\times
\left[\sum_{\boldsymbol{k}}V_{\boldsymbol{k}}^{g}M_{\boldsymbol{qk}}^{(1)}
\left(t,t_{1}\right)+\frac{1}{NS_{\boldsymbol{q}}}\left\langle
\rho_{\boldsymbol{q}}^{\star}\Omega^{\dagger}(t_{1})U_{1}^{\dagger}(t,t_{1})Q_{1
}g\right\rangle \right]\label{eq:g_t_simplified}\nonumber \\
 &  & +\int_{0}^{t}dt_{1}\underset{_{\boldsymbol{k^{\prime}}>\boldsymbol{p^{\prime}}}}{\sum_{\boldsymbol{k}>\boldsymbol{p}}}V_{0,\boldsymbol{k},\boldsymbol{p}}\left(t_{1}\right)V_{\boldsymbol{k^{\prime}},\boldsymbol{p^{\prime}}}^{g}\Phi_{\boldsymbol{kk^{\prime}}}(t,t_{1})\Phi_{\boldsymbol{pp^{\prime}}}(t,t_{1}).
\end{eqnarray}

We should pay attention to a possible issue with Eq.~\ref{eq:g_t_simplified}:
 if one is not cautious,
there is a risk that the uncontrolled
approximation
of (Orth1) may create spurious inhomogeneities that violate translational
invariance, even in cases where it should theoretically be obeyed. We shall ward
off this risk in a pragmatic way by making a
judicious choice for the transient density correlator and ensuring the
respect of fundamental physical principles in the final equations.  

\section{Severe approximations lead to a constitutive equation of the
White-Metzner class}

Armed with the projection scheme of the previous part, we are now
capable of studying the stress observable $\boldsymbol{\sigma}_{\boldsymbol{q}}$.
Direct application of Eq.~\ref{eq:g_t_simplified} gives

\begin{eqnarray}
\left\langle \boldsymbol{\sigma}_{\boldsymbol{q}}\right\rangle _{t}-\left\langle
\boldsymbol{\sigma}_{\boldsymbol{q}}\right\rangle _{0} & \approx &
-\int_{0}^{t}dt_{1}\sum_{\boldsymbol{q}}\left(\left\langle
\rho_{\boldsymbol{q}}\right\rangle _{t_{1}}-\left\langle
\rho_{\boldsymbol{q}}\right\rangle _{0}\right) \nonumber
 \\
 & &
\:\times \left[\sum_{\boldsymbol{k}}V_{\boldsymbol{k}}^{\boldsymbol{\sigma}_{
\boldsymbol {
q}}}M_{\boldsymbol{qk}}^{(1)}\left(t,t_{1}\right)+\frac{1}{NS_{\boldsymbol{q}}}
\left\langle
\rho_{\boldsymbol{q}}^{\star}\Omega^{\dagger}(t_{1})U_{1}^{\dagger}(t,t_{1})Q_{1
}\boldsymbol{\sigma}_{\boldsymbol{q}}\right\rangle \right]\nonumber \\
 &  & +\int_{0}^{t}dt_{1}\underset{_{\boldsymbol{k^{\prime}}>\boldsymbol{p^{\prime}}}}{\sum_{\boldsymbol{k}>\boldsymbol{p}}}V_{0,\boldsymbol{k},\boldsymbol{p}}\left(t_{1}\right)V_{\boldsymbol{k^{\prime}},\boldsymbol{p^{\prime}}}^{\boldsymbol{\sigma}_{\boldsymbol{q}}}\Phi_{\boldsymbol{kk^{\prime}}}(t,t_{1})\Phi_{\boldsymbol{pp^{\prime}}}(t,t_{1}).\label{eq:sigma_q_t}
\label{eq:stress_to_be_simpl}
 \end{eqnarray}

The usual MCT protocol involves the derivation of equations for the
transient density correlators $\Phi_{\boldsymbol{qp}}(t,t_{1})$,
with the help of the Zwanzig-Mori projection formalism. However, in
the presence of flow heterogeneities and in non-trivial geometry,
this would be a difficult task, which we bypass here:
bearing in mind our main goal, namely, a study of the visco-elastic
instability, we opt for a drastic simplification of the equations.
We shall thus perform very strong, and mostly uncontrolled (but explicitly
mentioned) approximations and we do
not expect them to preserve the quantitative details of the full
theory. Nevertheless, we aim to arrive at a tractable model that displays
shear-thinning and correctly captures the low-shear-rate and high-shear-rate
regime of the flow. This will allow us to test the phenomenology of our
extension of the framework to inhomogeneous
situations and flow instabilities.

\subsection{Reduction to a locally homogeneous flow \label{sec:red_loc_hom}}

To start with, we assume that the flow only moderately deviates from
homogeneity, so that the locally homogeneous flow approximation of
Section~\ref{sub:Leading-order-locally-homogeneou}
is valid. Under that assumption, the source term $\hat{S}_{\boldsymbol{p}}(t)$
for the density in Eq.~\ref{eq:density_equation} vanishes, and no
density inhomogeneity is created (effects such as shear-concentration
coupling \cite{Besseling2010,Jin2014} are thereby precluded). Thus,
we suppose $\left\langle \rho_{\boldsymbol{q}}\right\rangle _{t}\approx\left\langle \rho_{\boldsymbol{q}}\right\rangle _{0}\approx N\delta_{\boldsymbol{q},\boldsymbol{0}}$. 

Furthermore, if the velocity gradient varies little over the distance travelled
by a material volume before its microstructure relaxes, the intrinsic dynamics
of the local stress
$\boldsymbol{\sigma}(\boldsymbol{r},t)$ are governed by the history of the
\emph{local}
velocity gradient $\left\{ \boldsymbol{\kappa}(\boldsymbol{r},t^{\prime}),\, t^{\prime}<t\right\} $.
In other words, all non-local effects due to, \emph{e.g.}, stress waves emitted
by distant regions are
discarded. Within mesoscopic regions of the material, translational invariance
is obeyed, so that only one cross-coupling subsists in
the transient density correlator, \emph{viz.},
$\Phi_{\boldsymbol{kk^{\prime}}}(t,t_{1})=\Phi_{\boldsymbol{kk^{\prime}}}(t,t_{1
})\delta_{\boldsymbol{k},\boldsymbol{k^{\prime}}(t,t_{1})}$. Under these
assumptions, the vertices in Eq.~\ref{eq:stress_to_be_simpl} boil down to
\cite{Brader2012}

\begin{equation*}
V^\sigma_{\boldsymbol{k^\prime},\boldsymbol{p^\prime}} = \boldsymbol{k^\prime}
\boldsymbol{p^\prime} 
\frac{S^\prime_{\boldsymbol{k^\prime}}}{N^2 k^\prime S_{\boldsymbol{k^\prime}}
S_{\boldsymbol{p^\prime}}}
\delta_{\boldsymbol{k^\prime},-\boldsymbol{p^\prime}},
\end{equation*}

\begin{equation*}
V_{0,\boldsymbol{k},\boldsymbol{p}}(t_1)=
\boldsymbol{\kappa}(t_1):\boldsymbol{kp}
\frac{S^\prime_{\boldsymbol{k}}}{k}
\delta_{\boldsymbol{k},-\boldsymbol{p}}.
\end{equation*}
 
Thus,
we recover
the formula for homogeneous flow derived by Brader \emph{et al.}
\cite{Brader2009,Brader2012},
namely,

\begin{equation}
\boldsymbol{\sigma}(t)=-\int_{0}^{t}dt_{1}\sum_{\boldsymbol{k^\prime}}\,
A(\boldsymbol{k^\prime},t,t_{1})\left[\frac{\partial}{\partial
t_{1}}\left(\boldsymbol{k^\prime}\cdot\boldsymbol{B}(t,t_{1})\cdot\boldsymbol{
k^\prime}
\right)\right]\Phi_{\boldsymbol{k^\prime}(t,t_{1})}^{2}(t,t_{1})\boldsymbol{
k^\prime} \otimes\boldsymbol{k^\prime},\label{eq:ITT_MCT}
\end{equation}
where where we have introduced the Finger tensor $\boldsymbol{B}(t,t_1)\equiv
e_+^{\int_{t_1}^{t}ds \boldsymbol{\kappa}(s)}\cdot e_-^{\int_{t_1}^{t}ds
\boldsymbol{\kappa^\top}(s)} $ and
$A(\boldsymbol{k^\prime},t,t_{1}) \propto
\frac{S^\prime_{\boldsymbol{k^\prime}}S^\prime_{\boldsymbol{k^\prime(t,t_1)}}}{
k^\prime k^\prime(t, t_1)S^2_{ \boldsymbol{k^\prime}}} $ collects all relevant
equilibrium
properties of the material.

\subsection{Schematic approximation\label{sub:Schematic-approximation}}

To proceed, we follow the schematic approximation conducted in
Ref.~\cite{Brader2009} by dropping all existing wavevector dependences
(or, better said, by focusing on the most relevant wavevector) in
Eq.~\ref{eq:ITT_MCT}, \emph{viz.},
\begin{equation}
\boldsymbol{\sigma}(t)=-\upsilon_\sigma\int_{0}^{t}dt_{1}\frac{
\partial\boldsymbol{B}(t,t_{1})}{\partial
t_{1}}\Phi^{2}(t,t_{1}),\label{eq:sigma_pre_Lodge}
\end{equation}
 where $\upsilon_\sigma\approx k_{B}Tn$ (with $n$ the number density) sets the
scale of stress fluctuations. Partial integration yields
\begin{equation}
\boldsymbol{\sigma}(t)=\upsilon_\sigma\int_{0}^{t}dt_{1}\boldsymbol{B}(t,t_{1}
)\frac{\partial\Phi^{2}(t,t_{1})}{\partial t_{1}}.\label{sigma_Lodge}
\end{equation}

\subsection{Approximation of the transient density correlator}

At rest,  the thermal relaxation of density fluctuations takes
a time $\tau_{\alpha}$ that, according to MCT, diverges in the
ideal glassy phase. But the presence of a solvent flow distorts the
material structure and accelerates
this relaxation.  Arguably our crudest
assumption will now consist in
proposing a {}``phenomenological'' characteristic time scale $\tau$
for the long-run decay of the transient density correlator, which
takes this flow-induced relaxation into account: 
\begin{eqnarray}
\frac{\partial}{\partial t}\Phi(t,t_{1}) & = &
-\frac{\Phi(t,t_{1})}{2\tau\left[\boldsymbol{\dot{\epsilon}}\left(t\right)\right
]}\label{eq:approx_Phi}\\
\frac{\partial}{\partial t_{1}}\Phi(t,t_{1}) & = &
\frac{\Phi(t,t_{1})}{2\tau\left[\boldsymbol{\dot{\epsilon}}\left(t_{1}
\right)\right]}\label{eq:approx_Phi_2}\\
\text{with }\tau\left[\boldsymbol{\dot{\epsilon}}\right] & \equiv & \frac{\tau_{\alpha}}{1+2\alpha\tau_{\alpha}\sqrt{J_{2}\left(\boldsymbol{\dot{\epsilon}}\right)}},\label{eq:tau_def}
\end{eqnarray}
Here, $\boldsymbol{\dot{\epsilon}}\left(t\right)\equiv\frac{\boldsymbol{\kappa}\left(t\right)+\boldsymbol{\kappa}^{\top}\left(t\right)}{2}$
is the strain rate tensor, $J_{2}$ is the second tensorial invariant of
a deviatoric tensor, \emph{viz.}, $J_{2}\left(\boldsymbol{\dot{\epsilon}}\right)=\frac{1}{2}\dot{\epsilon}_{ij}\dot{\epsilon}_{ji}$
(this reduces to $J_{2}\left(\boldsymbol{\kappa}\right)=\frac{1}{4}\dot{\gamma}^{2}$
under simple shear), and $\alpha$ is a material parameter that quantifies
shear-thinning. More precisely, $\alpha^{-1}$ is an ``inverse yield
strain'' that describes how much strain
is required to erase the memory of the local structure \cite{Brader2009}. The
following limit
cases are
enlightening: at vanishing shear rate,
$\tau\left(\boldsymbol{\dot{\epsilon}}\right)$ tends to the quiescent
relaxation time $\tau_{\alpha}$, while at high shear rates one gets
$\tau\left(\boldsymbol{\dot{\epsilon}}\right) \approx (\alpha\dot{\gamma})^{-1}$
(under simple shear).

Equations~\ref{eq:approx_Phi}-\ref{eq:approx_Phi_2} lead to
\begin{equation}
\Phi^{2}(t,t_{1})=\mathrm{exp}\left(-\int_{t_{1}}^{t}\frac{ds}{\tau\left[
\boldsymbol{\dot{\epsilon}}\left(s\right)\right]}\right).\label{eq:phi2_approx}
\end{equation}

\subsection{Constitutive equation}

With these severe approximations in hand, we now differentiate Eq.~\ref{sigma_Lodge}
with respect to time $t$.

\begin{equation}
\frac{\partial}{\partial
t}\boldsymbol{\sigma}(t)=\frac{\upsilon_\sigma}{\tau\left[\boldsymbol{\dot{
\epsilon } } \left(t\right)\right]}\boldsymbol{\mathbb{I}}+\boldsymbol{\kappa}
(t)\cdot\boldsymbol{\sigma}(t)+\boldsymbol{\sigma}(t)\cdot\boldsymbol{\kappa}
(t)^{\top}-\frac{\boldsymbol{\sigma}(t)}{\tau\left[\boldsymbol{\dot{\epsilon}}
\left(t\right)\right]},\label{eq:sigma_der2}
\end{equation}
where $\boldsymbol{\mathbb{I}}$ is the identity matrix and we have used
the equality $\frac{\partial\Phi^{2}(t,t_{1})}{\partial
t_{1}}\Big|_{t_{1}=t}=\tau\left[\boldsymbol{\dot{\epsilon}}\left(t\right)\right]
^{-1}$
as well as $\frac{\partial\boldsymbol{B}(t,t_{1})}{\partial
t}=\boldsymbol{\kappa}(t)\cdot\boldsymbol{B}(t,t_{1})+\boldsymbol{B}(t,t_{1}
)\cdot\boldsymbol{\kappa}(t)^{\top}$.
Splitting $\boldsymbol{\sigma}$ into a quiescent
(\emph{i.e.}, $\boldsymbol{\kappa}=\boldsymbol{0}$) pressure part
$\upsilon_\sigma \mathbb{I}$ and a driven part
$\boldsymbol{\sigma}^{\mathrm{d}}$ in Eq.~\ref{eq:sigma_der2},
we arrive at  
\begin{equation}
\frac{\partial}{\partial
t}\boldsymbol{\sigma}^{\mathrm{d}}(t)=\upsilon_\sigma\left[\boldsymbol{\kappa}
\left(t\right)+\boldsymbol{\kappa}^{\top}\left(t\right)\right]+\boldsymbol{
\kappa}(t)\cdot
\boldsymbol{\sigma}^{\mathrm{d}}(t)+\boldsymbol{\sigma}^{\mathrm{d}}
(t)\cdot\boldsymbol {\kappa } (t)^ {
\top}-\frac{\boldsymbol{\sigma}^{\mathrm{d}}(t)}{\tau\left[\boldsymbol{\dot{
\epsilon}} \left(t\right)\right]}.
\label{eq:WM_intrinsic}
\end{equation}

  Finally, one should recall that Eq.~\ref{eq:WM_intrinsic} deals with
the intrinsic dynamics, \emph{i.e.}, the evolution in the (homogeneous)
auxiliary frame introduced in Eq.~\ref{eq:loc_hom_flow}. The full evolution of
the observable in the lab frame is recovered by applying Eq.~\ref{eq:dt_gs},
which restores the inhomogeneity advection term
$\boldsymbol{v}\cdot\nabla\boldsymbol{\sigma}$
found in Section~\ref{sub:Advection_term_gal}, responsible for the
advection of stress
fluctuation with the microstructure, hence with the flow. After dropping
the {}``d'' superscripts, we thus obtain

\begin{equation}
\frac{\mathfrak{D}}{\mathfrak{D}t}\boldsymbol{\sigma}(\boldsymbol{r},t)+\frac{
\boldsymbol{\sigma}(\boldsymbol{r},t)}{\tau\left[\boldsymbol{\dot{\epsilon}}
(\boldsymbol{r},t)\right]}=2\upsilon_\sigma\boldsymbol{\dot{\epsilon}}
(\boldsymbol { r } , t) , \label{eq:WM_const_eq}
\end{equation}
where the upper convected derivative (a.k.a. advected Truesdell derivative)
is defined by
\[
\frac{\mathfrak{D}\boldsymbol{\sigma}}{\mathfrak{D}t}\equiv\frac{\partial\boldsymbol{\sigma}}{\partial t}+\boldsymbol{v}\cdot\nabla\boldsymbol{\sigma}-\boldsymbol{\kappa}(t)\cdot\boldsymbol{\sigma}(t)-\boldsymbol{\sigma}(t)\cdot\boldsymbol{\kappa}(t)^{\top}.
\]
Remember that the final relaxation time depends on the intrinsic dynamics and
the shear rate and is given in Eq.~\ref{eq:tau_def}.

Interestingly, this relatively simple constitutive equation belongs
to a class of (mostly) phenomenological models initially put forward
in the context of polymer melt rheology by White and Metzner (WM)
\cite{White1963}, on the basis of symmetry considerations. It obeys the principles of locality,
causality and material objectivity.

In addition to the configuration-based stress $\boldsymbol{\sigma}$,
which is associated to the colloidal microstructure, we include a
Newtonian contribution to the stress accounting for viscous dissipation;
the latter can indeed become significant at large shear rates. The
total stress then reads
\begin{equation}
\boldsymbol{\Sigma}=\boldsymbol{\sigma}+2\eta_{s}\boldsymbol{\dot{\epsilon}}.\label{eq:total_stress_def}
\end{equation}
Note that this model has already been studied by Papenkort and Voigtmann
in the case of a channel flow \cite{Papenkort2013}, where however the advected derivative played no role.

\subsection{Bulk rheology and parameter fitting}

Within the WM-type model, we first consider the rheological properties
of the bulk shear flow, \emph{i.e.}, $\boldsymbol{\kappa}=\dot{\gamma}^{\star}\boldsymbol{e_{1}}\otimes\boldsymbol{e_{2}}$
, prior to any potential instability. From the constitutive equation, using
$\eta_{0}$ as a shorthand for
$\upsilon_\sigma\tau_{\alpha}$,
it is easy to show that

\begin{eqnarray}
\sigma_{11}^{\star} & = & 0\nonumber \\
\sigma_{12}^{\star} & = & \frac{\eta_{0}\dot{\gamma}^{\star}}{1+\alpha\tau_{\alpha}|\dot{\gamma}^{\star}|}\label{eq:bulk_sigma_12}\\
\sigma_{22}^{\star} & = & 2\eta_{0}\tau_{\alpha}\left(\frac{\dot{\gamma}^{\star}}{1+\alpha\tau_{\alpha}|\dot{\gamma}^{\star}|}\right)^{2}.\nonumber 
\end{eqnarray}

Turning to linear rheology, the storage and loss moduli associated
to the WM-type model are identical to those of a Maxwell model with
a Newtonian contribution, \emph{viz.}, 
\begin{eqnarray*}
G^{\prime}(\omega) & = & \frac{\eta_{0}\tau_{\alpha}\omega^{2}}{1+\omega^{2}\tau_{\alpha}^{2}}\\
G^{\prime\prime}(\omega) & = & \frac{\eta_{0}\omega}{1+\omega^{2}\tau_{\alpha}^{2}}+\eta_{s}\omega.
\end{eqnarray*}

To fit the model parameters $\upsilon_\sigma$, $\tau_{\alpha}$, $\alpha$, and
$\eta_{s}$, we compare the $\sigma_{12}^{\star}\left(\dot{\gamma}^{\star}\right)$-flow
curve and, with less emphasis, the storage and loss moduli to experimental
measurements by Siebenb\"urger \emph{et al.} on suspensions of
$\sim100\,\mathrm{nm}$-large
colloids with a thermosensitive PNIPAM shell (which affords a sensitive
control of the effective volume fraction through the tuning of the
temperature) \cite{Siebenburger2009}. In Ref.~\cite{Siebenburger2009},
a schematic version of the MCT equations was shown to
provide excellent agreement with both the oscillatory shear and the steady
shear measurements, for several effective volume fractions across
the glass transition, while  the solutions of
the \emph{microscopic} MCT equations were tested in
Ref.~\cite{Amann2015nonlinear} and
yielded a consistent first principles description.  

Figure~\ref{fig:bulk_rheology_fits} presents the flow curve fits
obtained with our much cruder (but also much more tractable) constitutive
equation; there is no doubt that the quality of the fits has suffered
from our strong approximations: this confirms that the memory effects
encoded in $\Phi(t,t_{1})$ are more subtle than our simple Ansatz
of Eq.~\ref{eq:tau_def}. Nevertheless, the downward bending of the
stress at low shear rates, which originates from thermal relaxation,
and the strong shear-thinning effects, as well as the viscous behaviour
at high shear rates are reasonably well captured. Let us also mention
that our fitted model parameters (see Table~\ref{tab:Model-parameters})
have values comparable to the corresponding quantities in Ref.~\cite{Siebenburger2009};
note, for instance, that $\tau_{\alpha}$ increases dramatically with
$\phi_{\mathrm{eff}}$ and that $\alpha$ is roughly of the order
of the inverse yield strain.

We bestowed less importance to the linear rheology moduli $G^{\prime}$
and $G^{\prime\prime}$ in fitting the model parameters, because the
rest of the paper deals with steady shear; still, Fig.~\ref{fig:lin_rheology_fits}
shows that the experimentally observed trends are also present in
the model, although there is no \emph{quantitative} agreement.
The origin of the deviations is the broad distribution of relaxation times neglected in the White Metzner model but contained in the MCT description.

\begin{center}
\begin{figure}
\begin{centering}
\subfloat[$\phi_{\mathrm{eff}}=0.519$]
{\includegraphics[width=5.5cm]{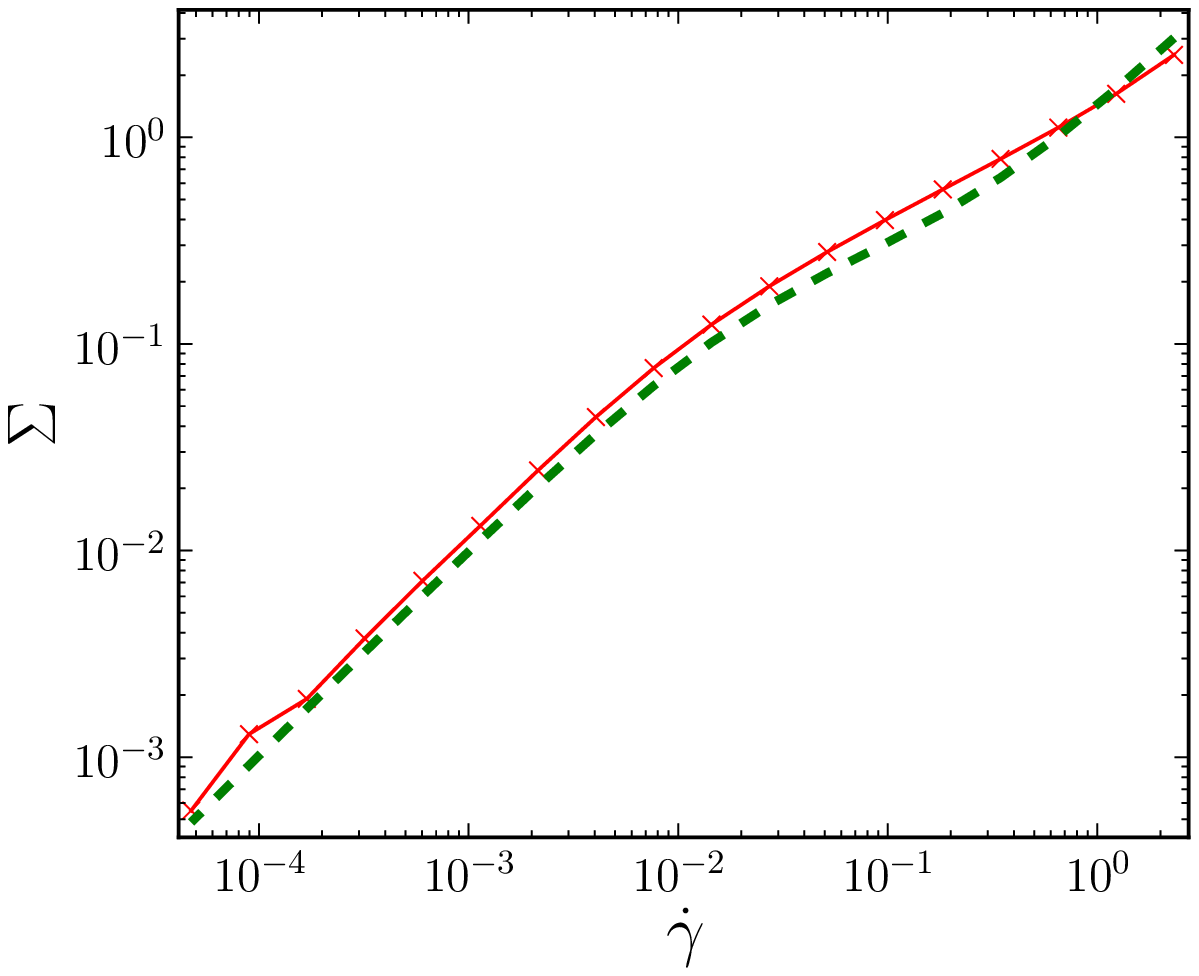}}
\hspace{1cm}\subfloat[$\phi_{\mathrm{eff}}=0.600$]{
\includegraphics[width=5.5cm]{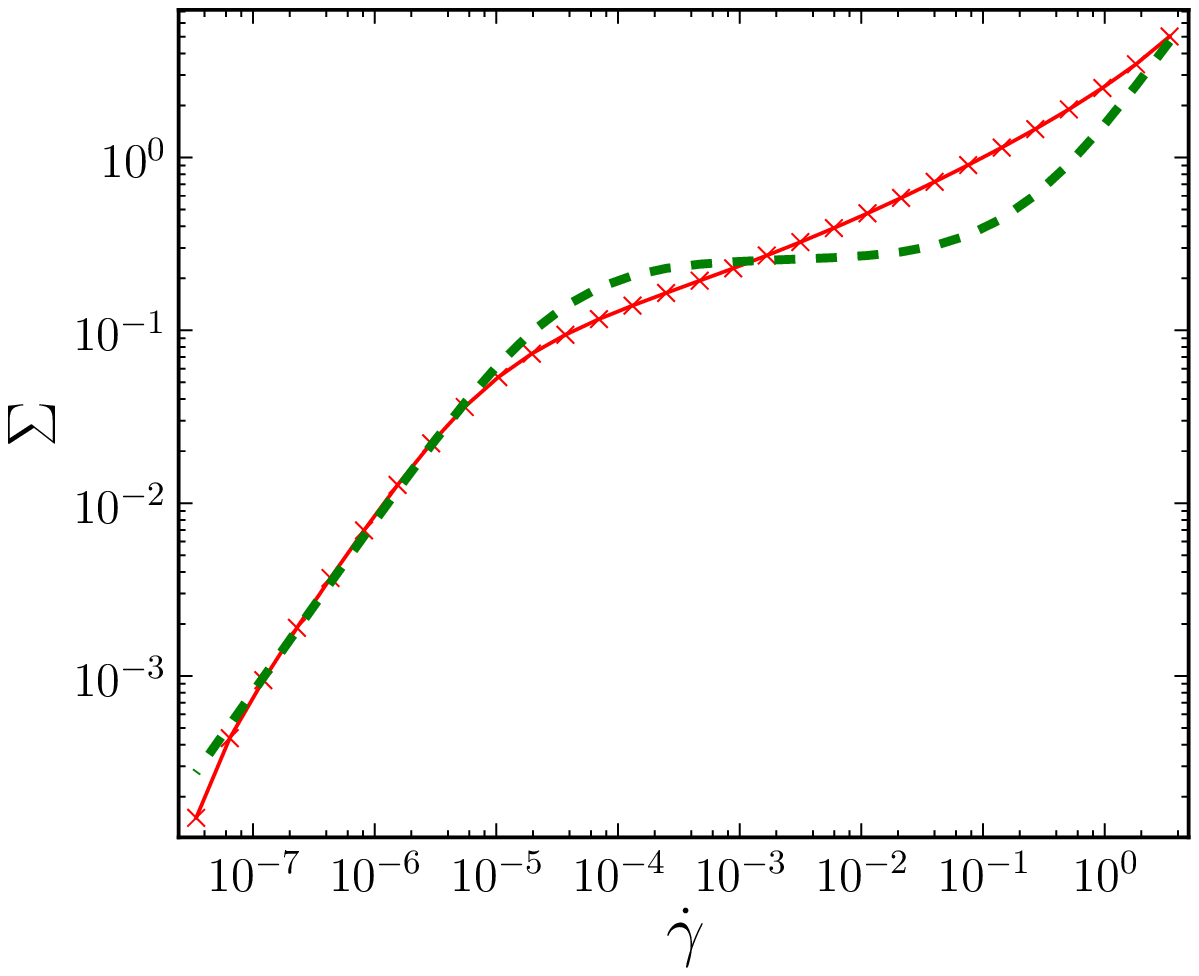}

}
\par\end{centering}

\begin{centering}
\subfloat[$\phi_{\mathrm{eff}}=0.626$]{
\includegraphics[width=5.5cm]{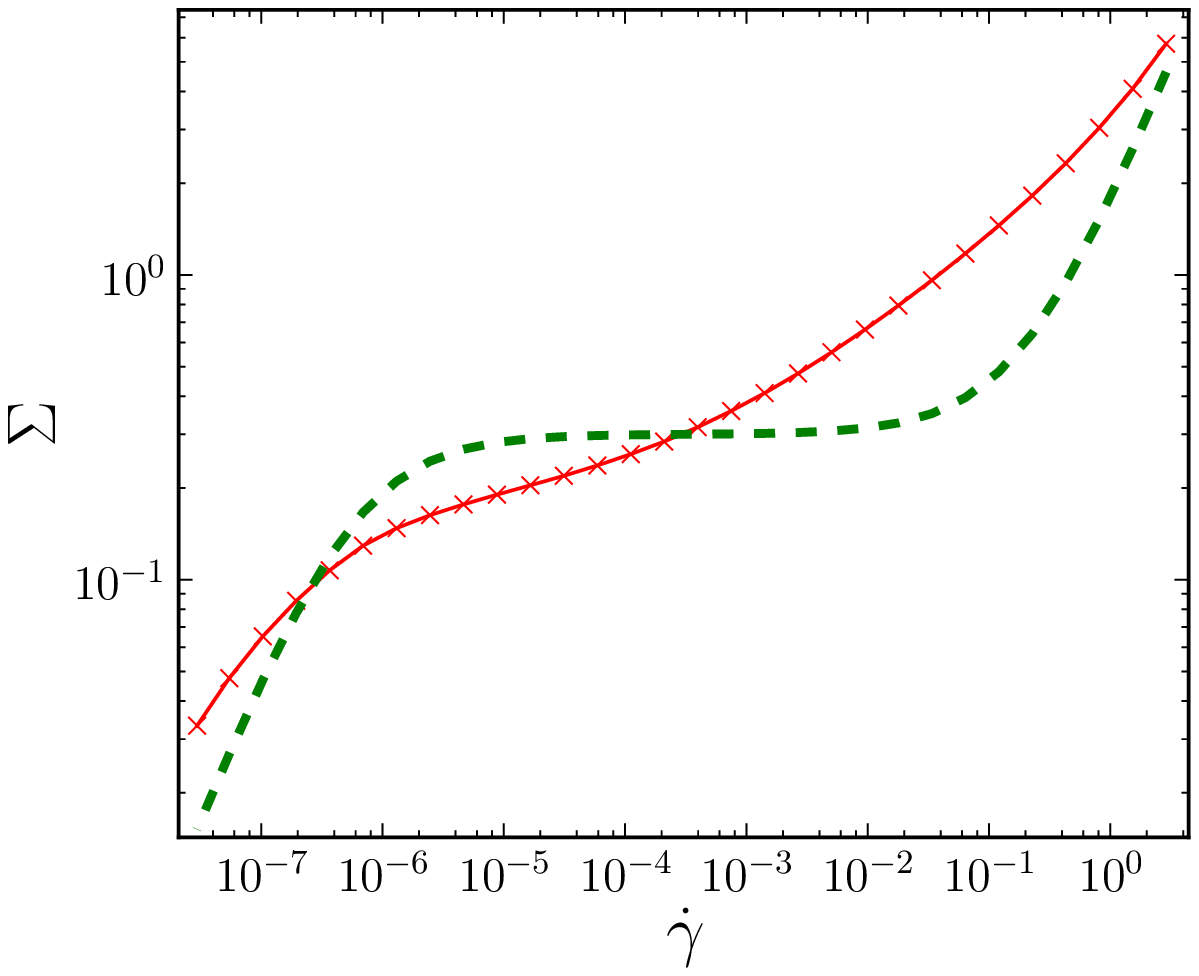}
}
\hspace{1cm}\subfloat[$\phi_{\mathrm{eff}}=0.639$]{
\includegraphics[width=5.5cm]{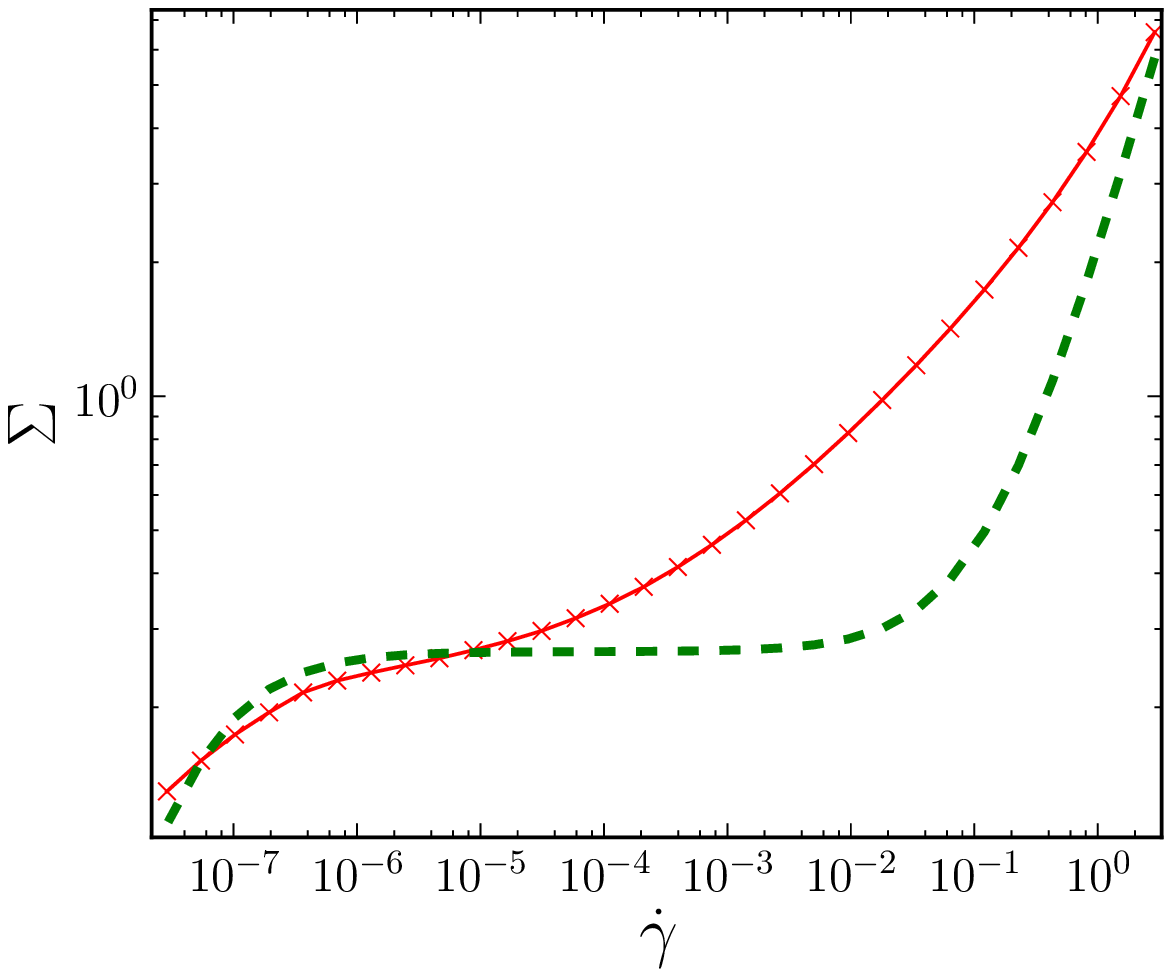}

}
\par\end{centering}

\begin{centering}
\subfloat[$\phi_{\mathrm{eff}}=0.641$]{
\includegraphics[width=5.5cm]{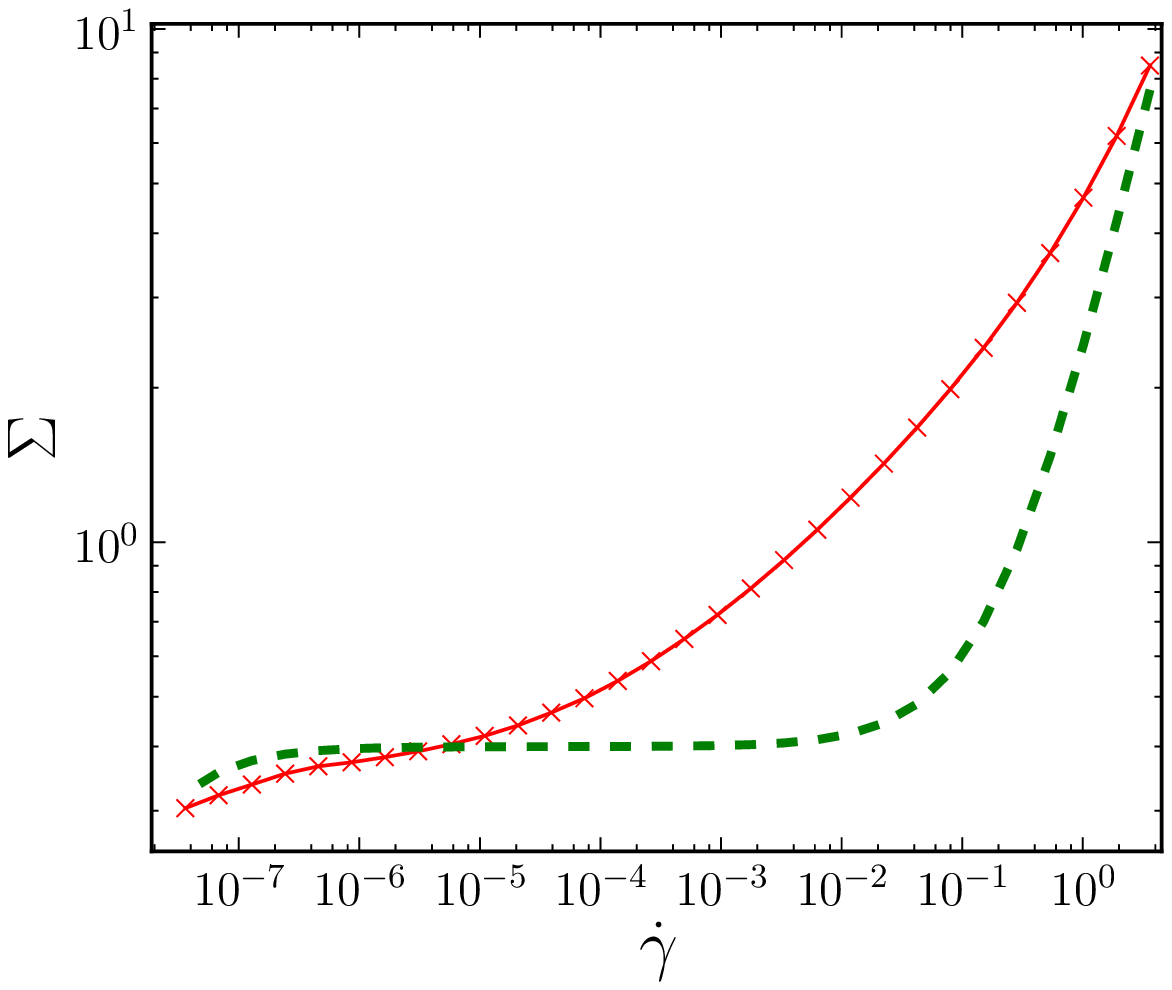}

}
\par\end{centering}

\caption{\label{fig:bulk_rheology_fits}Flow curves for different effective
volume fractions. (\emph{Red lines with crosses}) experimental measurements
from Ref.~{\cite{Siebenburger2009}}; (\emph{dashed
green lines}) fits with the WM-type model (see text); parameters are listed in Table I.\protect \\
Stresses are given in units of $\frac{R_{h}^{3}}{k_{B}T}$, where
$R_{h}$ is the hydrodynamic radius of the colloids, and shear rates
are non-dimensionalised with $\frac{6\pi\eta_{\mathrm{solv}}R_{h}^{3}}{k_{B}T}$,
with $\eta_{\mathrm{solv}}$ the solvent viscosity. }

\end{figure}

\par\end{center}

\begin{center}
\begin{figure}
\begin{centering}
\subfloat[$\phi_{\mathrm{eff}}=0.519$]{\includegraphics[width=5.5cm]{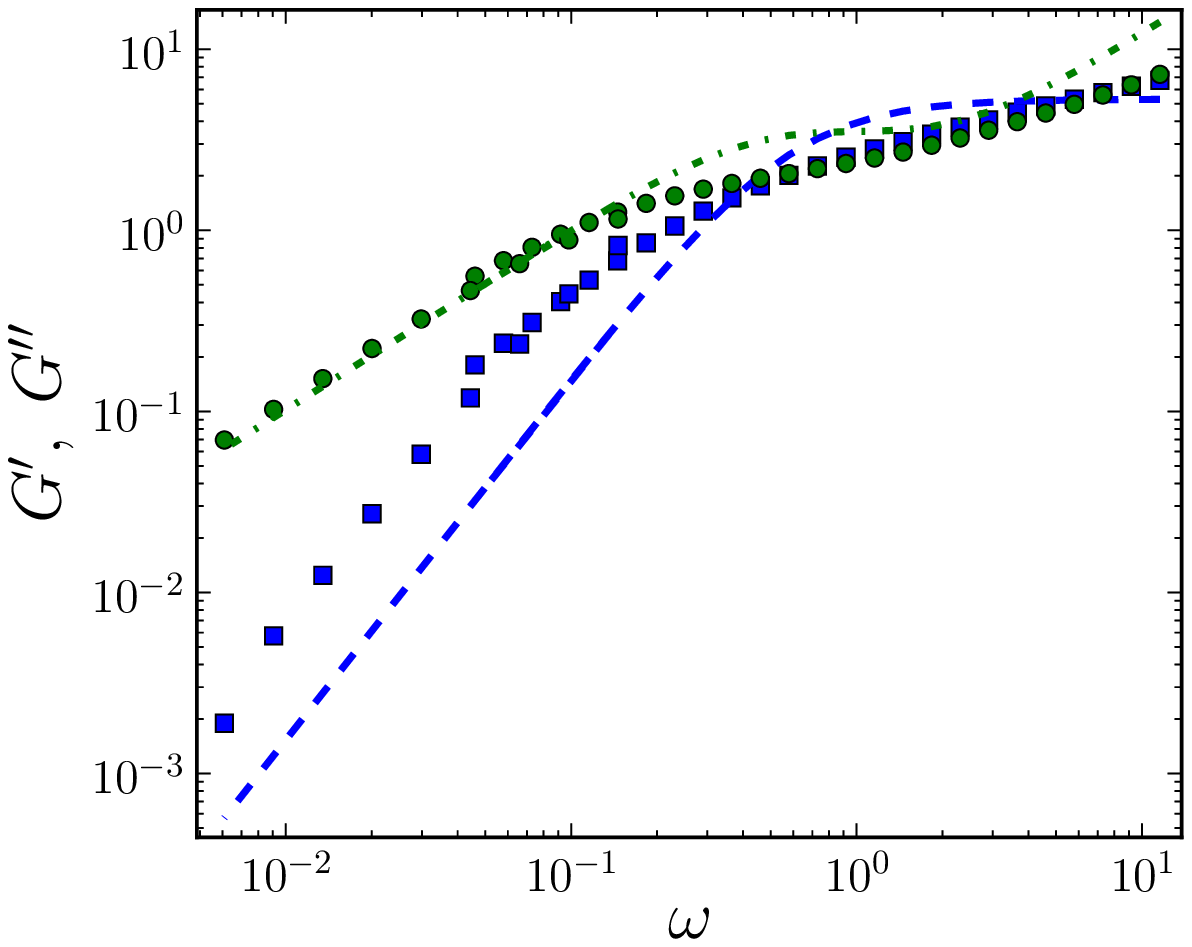}

}\hspace{1cm}\subfloat[$\phi_{\mathrm{eff}}=0.600$]{\includegraphics[width=5.5cm
]{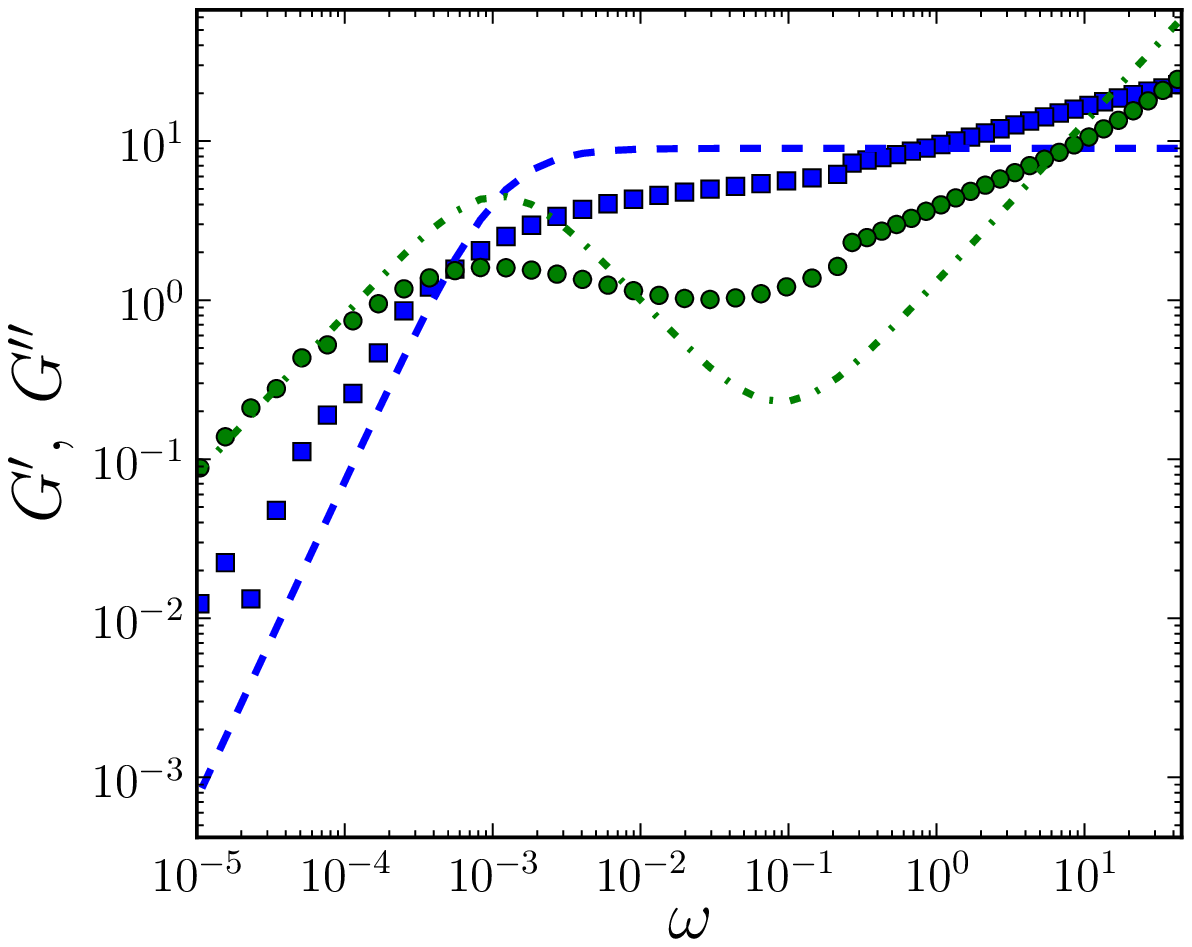}

}
\par\end{centering}

\begin{centering}
\subfloat[$\phi_{\mathrm{eff}}=0.626$]{\includegraphics[width=5.5cm]{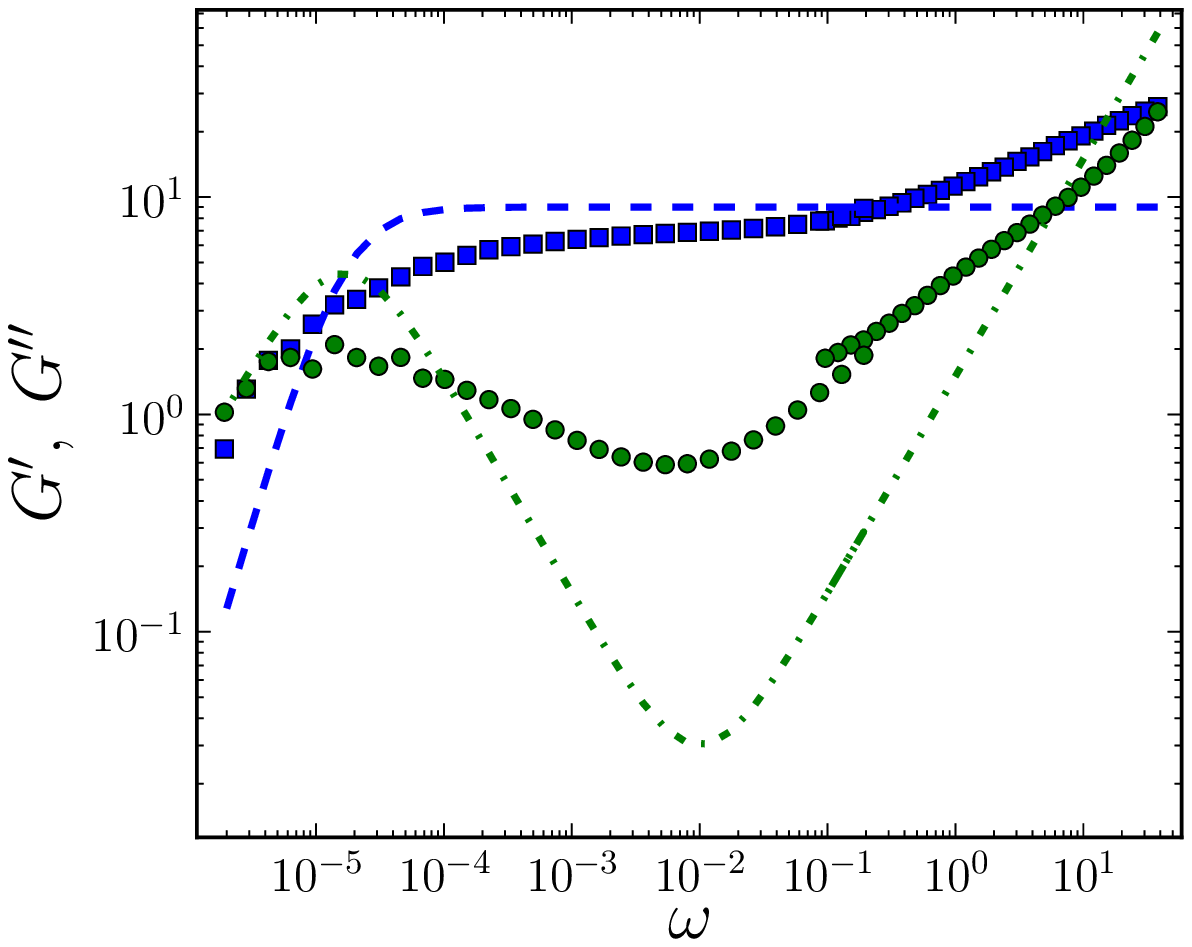}

}\hspace{1cm}\subfloat[$\phi_{\mathrm{eff}}=0.639$]{\includegraphics[width=5.5cm
]{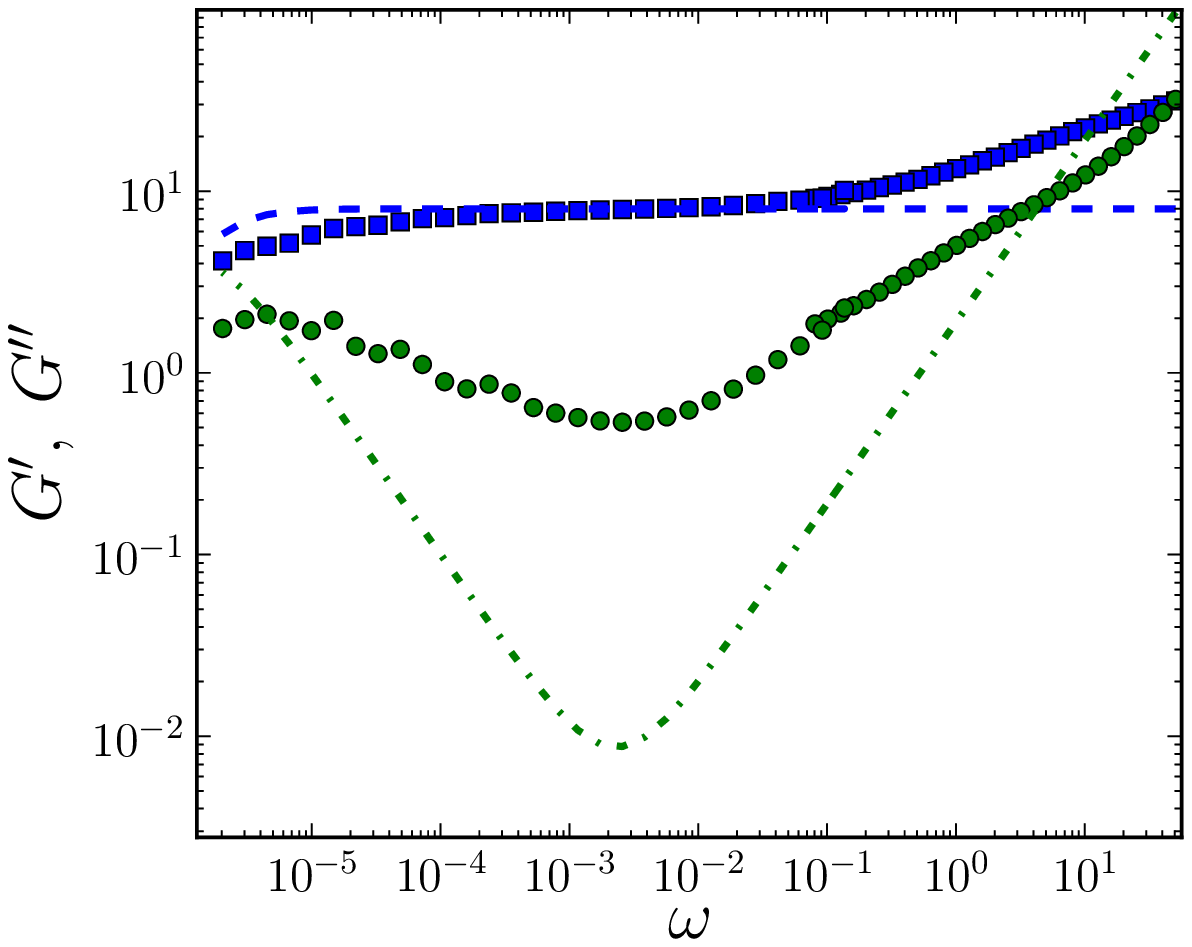}

}
\par\end{centering}

\begin{centering}
\subfloat[$\phi_{\mathrm{eff}}=0.641$]{\includegraphics[width=5.5cm]{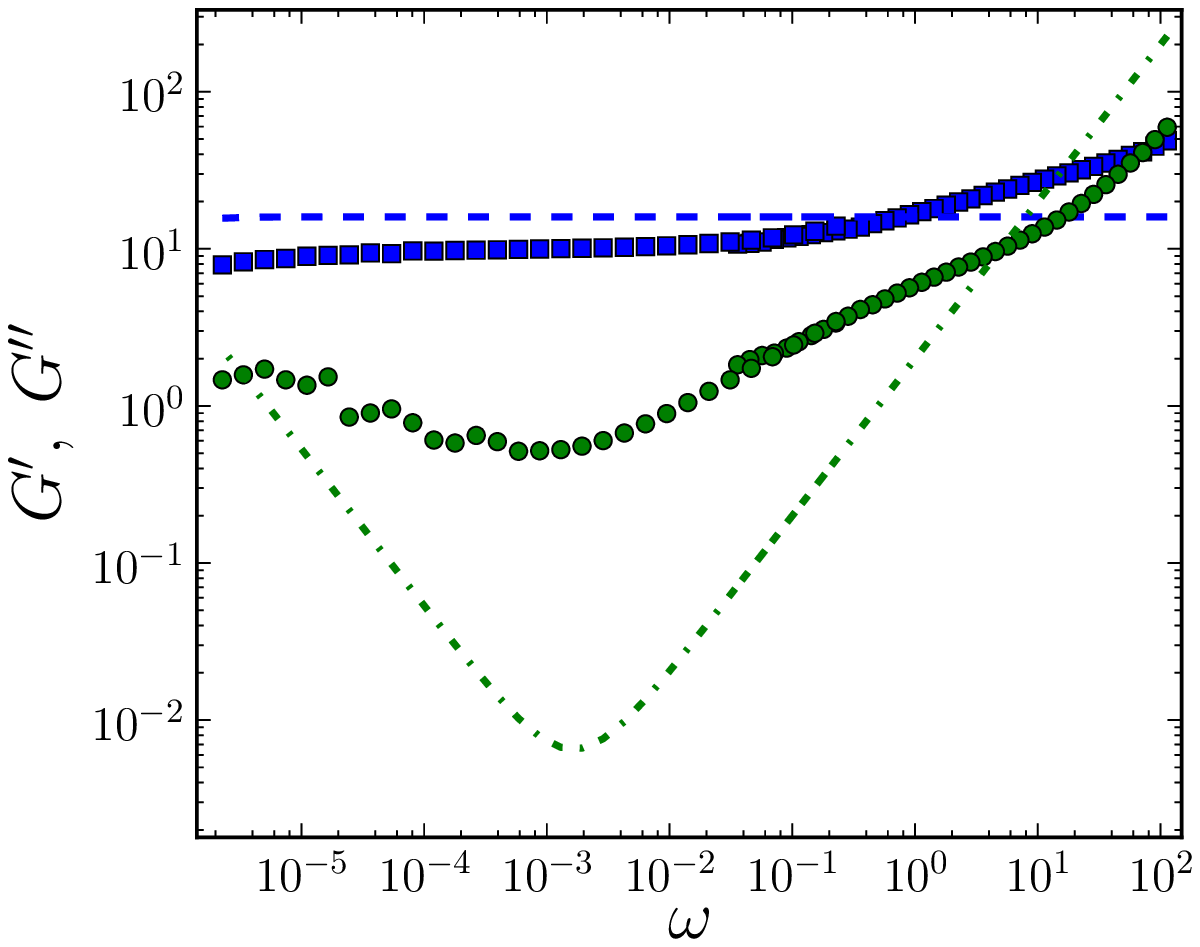}

}
\par\end{centering}

\caption{\label{fig:lin_rheology_fits}Linear rheology measurements for different
effective volume fractions. (\emph{Blue squares}) experimental storage
moduli $G^{\prime}(\omega)$ from Ref.~{\cite{Siebenburger2009}};
(\emph{dashed blue line}) fit with the WM-type model (see text). (\emph{Green
dots}) experimental loss moduli $G^{\prime\prime}(\omega)$; (\emph{dashed
green line}) fit with the WM-type model. The fitting parameters, as
well as the stress and frequency units, are identical to those used
in Fig.~\ref{fig:bulk_rheology_fits}.}
\end{figure}

\par\end{center}

\begin{center}
\begin{table}
\begin{centering}
\begin{tabular}{|c|c|c|c|c|}
\hline 
$\phi_{\mathrm{eff}}$ & $\tau_{\alpha}$ & $\alpha$ & $\eta_{0}$ & $\eta_{s}$\tabularnewline
\hline 
\hline 
0.519  & $1.7$ & $22$ & $9.0$ & $1.2$\tabularnewline
\hline 
0.600  & $900$ & $35$ & $8.1\cdot10^{3}$ & $1.3$\tabularnewline
\hline 
0.626  & $6.0\cdot10^{4}$ & $30$ & $5.4\cdot10^{5}$ & $1.5$\tabularnewline
\hline 
0.639  & $8.0\cdot10^{5}$ & $30$ & $6.4\cdot10^{6}$ & $1.9$\tabularnewline
\hline 
0.641  & $3.0\cdot10^{6}$ & $40$ & $4.8\cdot10^{7}$ & $2.0$\tabularnewline
\hline 
\end{tabular}
\par\end{centering}

\caption{\label{tab:Model-parameters}Model parameters used to fit the experimental
flow curves in Fig.~\ref{fig:bulk_rheology_fits} and linear spectra in Fig.~\ref{fig:lin_rheology_fits} .\protect \\
As in Fig.~\ref{fig:bulk_rheology_fits}, stresses are in units of
$\frac{R_{h}^{3}}{k_{B}T}$ and times in units of $\frac{k_{B}T}{6\pi\eta_{\mathrm{solv}}R_{h}^{3}}$.}
\end{table}

\par\end{center}

\subsection{Base flow in Taylor-Couette geometry\label{sub:Base-flow-in}}

Now, we focus on the base flow in a curved geometry, and more specifically
in the Taylor-Couette cell used by Siebenb\"urger and colleagues in Ref.~\cite{Siebenburger2009}. In such
a rheometer, the fluid flows in the annular region between two co-axial
cylinders, of radii $r_{i}=13.33\,\mathrm{mm}$ and $r_{o}=14.46\,\mathrm{mm}$
(relative gap width $\epsilon=0.085$), due to the rotation of the
inner one. This geometry will be kept fixed for the rest of the paper.

Turning to the determination of the flow, we note that, unlike in
Section~\ref{sec:Theoretical-framework-for}, the velocity profile
$\boldsymbol{v}(\boldsymbol{r})$ is no longer prescribed. Therefore,
we need to close the equations by complementing
the WM-type constitutive equation, (Eq.~\ref{eq:WM_const_eq}) with
the inertialess momentum conservation equation,
\begin{equation}
0=\nabla\cdot\boldsymbol{\Sigma}-\nabla p,\label{eq:Stokes}
\end{equation}
where $p$ is the pressure, and with the postulate of incompressibility,
\begin{equation}
0=\nabla\cdot\boldsymbol{v}.\label{eq:incompressibility}
\end{equation}

\begin{figure}
\begin{centering}
\includegraphics[width=6cm]{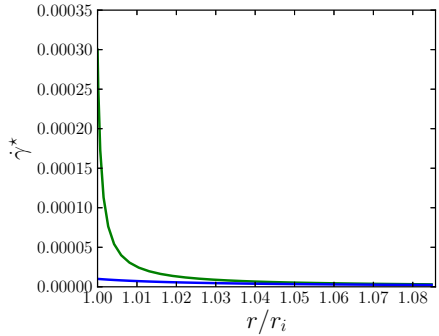}
\par\end{centering}

\caption{\label{fig:Velocity-profile-across}Shear rate profile of the base flow
in the Taylor-Couette cell, with model parameters corresponding to
$\phi_{\mathrm{eff}}=0.626$,
for imposed (non-dimensional) shear rates at the rotor
(\emph{blue}) $\dot{\gamma}(r_{i})=10^{-5}$ and 
 (\emph{green}) $\dot{\gamma}(r_{i})=3\cdot10^{-4}$.}

\end{figure}

In the considered geometry, the base flow is purely azimuthal and
has no dependence on $\theta$ or $z$, in cylindrical coordinates.
It follows from Eq.~\ref{eq:Stokes} that the total shear stress
must satisfy
\[
\Sigma\left(r\right)=\Sigma\left(r_{i}\right)\frac{r_{i}^{2}}{r^{2}}.
\]
 Inversion of Eqs.~\ref{eq:total_stress_def} and \ref{eq:bulk_sigma_12}
yields, for $\dot{\gamma}(r_{i})>0$,

\[
\dot{\gamma}^{\star}(r)=\frac{1}{2\alpha\eta_{s}}\left[\alpha\tau_{\alpha}
\Sigma\left(r\right)-\eta_{0}-\eta_{s}+\sqrt{\left(\eta_{0}+\eta_{s}-\alpha\tau_
{\alpha}\Sigma\left(r\right)\right)^{2}+4\eta_{s}\alpha\tau_{\alpha}
\Sigma\left(r\right)}\right],
\]
a profile of which is plotted in
Fig.~\ref{fig:Velocity-profile-across} for a particular applied shear rate.
Since $\dot{\gamma}^{\star}(r)\equiv
v_{\theta}^{\star\,\prime}-\frac{v_{\theta}^{\star}}{r}$,
we can solve numerically for the velocity profile $v_{\theta}(r)$.
It is perhaps worth indicating that an analytical solution exists
for $\eta_{s}=0$:

\begin{equation}
v_{\theta}^{\star}(r)=\frac{r}{\alpha\tau_{\alpha}}\ln\sqrt{\frac{
\upsilon_\sigma\alpha^ { -1 }
-\Sigma(r)}{\upsilon_\sigma\alpha^{-1}-\Sigma(r)}}.
\label{eq:analytical_inviscid_velocity}
\end{equation}
Incidentally, note that the quality of the approximation $\eta_{s}=0$
is not fixed by the inequality $\eta_{s}\ll\eta_{0}$, but by the
more stringent condition
$\eta_{s}\ll\frac{\eta_{0}}{1+\alpha\tau_{\alpha}|\dot{\gamma}^{\star}|}$.

\section{Shear-Thinning suppresses the visco-elastic instability}

\subsection{The visco-elastic instability}

For almost one century, it has been known that the inertial flow of
Newtonian liquids is prone to a centrifugal instability at large Reynolds
numbers (or, more precisely, at large Taylor numbers), whereby counter-rotating
vortices, known as {}``Taylor vortices'', appear and break the full
cylindrical symmetry of the base flow \cite{Taylor1923,Fardin2014hydrogen}.
Shear-thinning fluids are also prone to this instability \cite{Coronado2004},
which may even be enhanced, owing to shear-thinning, for polymeric
solutions \cite{Larson1989}.

This type of instability is driven by inertia and opposed
by viscosity. In dense colloidal suspensions, the vanishing Reynolds
number precludes it. Yet, there exists a distinct
type of instability, which does not require inertia: the so called
visco-elastic instability was first analysed by Muller, Larson and
Shaqfeh \cite{Muller1989,Larson1990,Shaqfeh1992} and has notably
been observed in polymer solutions \cite{Muller1989}
and (semi-dilute) worm-like micellar solutions~\cite{Fardin2012a},
in Taylor-Couette, cone-and-plate and parallel-plate rheometers (for
a review, see Ref. \cite{Shaqfeh1996}). To grasp the importance of
this finding, one should remember that, for decades, scientists have
measured the rheological properties of polymer solutions and polymer
melts in such setups, under the assumption of a well-defined (base)
flow.

The precise mechanism driving the instability is still, to some extent,
unsettled, but it is clear that curved streamlines and material elasticity
(\emph{i.e.}, a finite structural relaxation time) are vital for the
(linear) instability to develop. Indeed, because of the curvature
in, say,  a cylindrical setup, the normal stress $\Sigma_{\theta\theta}$
is coupled to the radial velocity component; this can lead to a positive
feedback mechanism, whereby a spontaneous radial velocity perturbation
alters $\Sigma_{\theta\theta}$, which further amplifies the perturbation
\cite{Shaqfeh1996}. To assess the effect of the material's properties
on the instability, Larson used a model featuring a distribution
of relaxation times and an adjustable shear-thinning exponent, and showed that
shear-thinning tends to stabilise the flow, by shifting the unstable
region (in parameter space) to larger applied shear rates \cite{Larson1994}.
Nevertheless, the author noted that {}``an elastic instability can
occur even in highly shear-thinning entangled polymer solutions'',
in the light of his calculations and experiments. Clearly, we are
interested in knowing whether this also holds true for highly concentrated
colloidal suspensions: Does the dramatic shear-thinning behaviour
of these materials allow for a visco-elastic instability within the
experimentally probed range of shear rates, according to the model
introduced in the previous section?

\subsection{Pseudo-spectral method for linear stability analysis}

To address this question, we analyse the stability of the base flow
in Taylor-Couette geometry (see Section~\ref{sub:Base-flow-in})
with respect to linear perturbations. Although many studies have focused
only on axisymmetric instabilities, non-axisymmetric modes were shown
to be even more unstable in a number of cases \cite{Avgousti1993,Nicolas2012},
so we investigate three-dimensional perturbations, \emph{i.e.}, perturbations
$\delta\phi$ with a spatial dependence not only on $r$ and $z$,
but also on $\theta$, \emph{viz.}, 
\[
\delta\phi(r,\theta,z,t)\equiv\left(\begin{array}{cccccccccc}
\delta\sigma_{rr} & \delta\sigma_{r\theta} & \delta\sigma_{rz} & \delta\sigma_{\theta\theta} & \delta\sigma_{\theta z} & \delta\sigma_{zz} & \delta v_{r} & \delta v_{\theta} & \delta v_{z} & \delta p\end{array}\right)^{\top}.
\]
We resort to a pseudo-spectral method. 

(i) We start by linearising the equations of the problem, comprising
the six constitutive equations (Eqs.~\ref{eq:WM_const_eq}), the
three momentum conservation equations (Eqs.~\ref{eq:Stokes}), and
the incompressibility postulate (Eq.~\ref{eq:incompressibility}),
around the base flow so as to obtain a system of linear equations
of the form
\[
\underset{\mathcal{A}}{\underbrace{\left(\begin{array}{ccccccc}
\frac{\partial}{\partial t}\\
 & \ddots\\
 &  & \frac{\partial}{\partial t}\\
 &  &  & 0\\
 &  &  &  & 0\\
 &  &  &  &  & 0\\
 &  &  &  &  &  & 0
\end{array}\right)}}\delta\phi=\mathcal{L}^{\star}\delta\phi.
\]
Then we transform the endomorphisms $\mathcal{A}$
and $\mathcal{L}^{\star}$ into matrices,
\emph{viz.}, $\mathcal{A}\rightarrow{\bf A}$ and $\mathcal{L}^{\star}\rightarrow{\bf L}^{\star}$,
with the following generic procedure: 

(ii) $\delta\phi$ is Fourier-transformed both in the azimuthal direction
(wavenumber $m$) and in the axial direction (wavenumber $k$), \emph{viz.}, 

\[
\delta\phi(r,\theta,z,t)=\sum_{m=-\infty}^{\infty}\sum_{k\in\nicefrac{2\pi}{L_{z}}\mathbb{Z}}\delta\phi(r,m,k,t)\, e^{im\theta}e^{ikz},
\]

(iii) $\delta\phi$ is discretised along the radial coordinate $r$
by sampling its values at the Chebyshev\textendash{}Gauss\textendash{}Lobatto
points $r_{n}\equiv1.0+\frac{\epsilon}{2}\left[1+\mathrm{cos}\left(\pi\frac{n}{M}\right)\right],\,0\leqslant n<M$,
for a given $M\in\mathbb{N}^{\star}$. We will typically use $M\approx2^{6}$
interpolation points across the gap. Radial derivatives $\partial_{r}\delta\phi$
are then written as matrix-vector products of the form $D_{r}\cdot\delta\phi$
\cite{Peyret2002}.

(iv) $\delta\phi$ is Laplace-transformed with respect to time, with
Laplace coordinate $s$,\emph{ viz.}, $\delta\phi(r,m,k,s)=\int e^{st}\delta\phi(r,m,k,t)dt$.

Eventually, one obtains the following generalised eigenvalue problem,
\begin{equation}
{\bf A}\delta\phi(r,m,k,s)=s{\bf L}^{\star}\delta\phi(r,m,k,s),\label{eq:genEvPb}
\end{equation}
where ${\bf A}$ and ${\bf L}^{\star}$ are $10M\times10M$ matrices,
a few lines of which are subsequently substituted for the implementation
of the (no-slip) boundary conditions on the velocity. 

Let $s^{\star}$ be the maximal growth rate, \emph{i.e., }the real
part of the maximal eigenvalue\footnote{
Due to the discretisation, some spurious eigenvalues may pop up in
the generalised eigenvalue problem (Eq.~\ref{eq:genEvPb}), but they
can easily be eliminated because, unlike their physical counterparts,
they vary with the number of discretisation points.%
} of Eq.~\ref{eq:genEvPb} over all possible wavenumbers $m$ and
$k$. Then the base flow is (linearly) stable if, and only if, $s^{\star}<0$.

\subsection{Linearised equations}

The linearised constitutive equations read:
\begin{eqnarray*}
\frac{\partial\delta\sigma_{rr}}{\partial t} & = &
\left(\frac{-1}{\tau^{\star}}-\frac{im}{r}v_{\theta}^{\star}\right)\delta\sigma_
{rr}+\left(2\sigma_{r\theta}^{\star}\frac{im}{r}+2\upsilon_\sigma\partial_{r}
\right)\delta v_{r}\\
\frac{\partial\delta\sigma_{r\theta}}{\partial t} & = &
\dot{\gamma}^{\star}\delta\sigma_{rr}+\left(\frac{-1}{\tau^{\star}}-\frac{im}{r}
v_{\theta}^{\star}\right)\delta\sigma_{r\theta}+\left[\sigma_{r\theta}^{\star}
\frac{im}{r}+\upsilon_\sigma\left(\partial_{r}-\frac{1}{r}\right)-\alpha\sigma_{
r\theta}^{\star}\left(\partial_{r}-\frac{1}{r}\right)\right]\delta v_{\theta}\\
 &  &
+\left[\sigma_{r\theta}^{\star}\left(\partial_{r}+\frac{1}{r}\right)-\partial_{r
}\sigma_{r\theta}^{\star}+\sigma_{\theta\theta}^{\star}\frac{im}{r}
+\upsilon_\sigma\frac{im}{r}-\alpha\sigma_{r\theta}^{\star}\frac{im}{r}\right]
\delta v_{r}\\
\frac{\partial\delta\sigma_{rz}}{\partial t} & = &
\left(\frac{-1}{\tau^{\star}}-\frac{im}{r}v_{\theta}^{\star}\right)\delta\sigma_
{rz}+ik\upsilon_\sigma\delta
v_{r}+\left(\sigma_{r\theta}^{\star}\frac{im}{r}+\upsilon_\sigma\partial_{r}
\right)\delta v_{z}\\
\frac{\partial\delta\sigma_{\theta\theta}}{\partial t} & = &
2\dot{\gamma}^{\star}\delta\sigma_{r\theta}+\left(\frac{-1}{\tau^{\star}}-\frac{
im}{r}v_{\theta}^{\star}\right)\delta\sigma_{\theta\theta}+\left(\frac{2\sigma_{
\theta\theta}^{\star}}{r}-\partial_{r}\sigma_{\theta\theta}^{\star}+\frac{
2\upsilon_\sigma}{r}-\alpha\sigma_{\theta\theta}^{\star}\frac{im}{r}
\right)\delta v_{r}\\
 &  &
+\left[2\sigma_{r\theta}^{\star}(\partial_{r}-\nicefrac{1}{r})+2\sigma_{
\theta\theta}^{\star}\frac{im}{r}+2\upsilon_\sigma\frac{im}{r}-\alpha\sigma_{
\theta\theta}^{\star}\left(\partial_{r}-\frac{1}{r}\right)\right]\delta
v_{\theta}\\
\frac{\partial\delta\sigma_{\theta z}}{\partial t} & = &
\dot{\gamma}^{\star}\delta\sigma_{rz}+\left(\frac{-1}{\tau^{\star}}-\frac{im}{r}
v_{\theta}^{\star}\right)\delta\sigma_{\theta z}+ik\upsilon_\sigma\delta
v_{\theta}+\left(\upsilon_\sigma\frac{im}{r}+\sigma_{\theta\theta}^{\star}\frac{
im}{r} +\sigma_{r\theta}^{\star}\partial_{r}\right)\delta v_{z}\\
\frac{\partial\delta\sigma_{zz}}{\partial t} & = &
\left(\frac{-1}{\tau^{\star}}-\frac{im}{r}v_{\theta}^{\star}\right)\delta\sigma_
{zz}+2ik\upsilon_\sigma\delta v_{z},
\end{eqnarray*}
where $\tau^{\star}\equiv\frac{\tau_{\alpha}}{1+\alpha\tau_{\alpha}\dot{\gamma}^{\star}}$,
while the linearised momentum conservation equations are
\begin{eqnarray*}
0 & = & \left(\partial_{r}+\frac{1}{r}\right)\delta\sigma_{rr}+\frac{im}{r}\delta\sigma_{r\theta}+ik\delta\sigma_{rz}-\frac{1}{r}\delta\sigma_{\theta\theta}-\partial_{r}\delta p\\
 &  & +\eta_{s}\left[\partial_{r}^{2}+\frac{1}{r}\partial_{r}-\frac{1}{r^{2}}-\frac{m^{2}}{r^{2}}-k^{2}\right]\delta v_{r}-2\eta_{s}\frac{im}{r^{2}}\delta v_{\theta}\\
0 & = & \left(\partial_{r}+\frac{2}{r}\right)\delta\sigma_{r\theta}+\frac{im}{r}\delta\sigma_{\theta\theta}+ik\delta\sigma_{\theta z}-\frac{im}{r}\delta p\\
 &  & +\eta_{s}\left[\partial_{r}^{2}+\frac{1}{r}\partial_{r}-\frac{m^{2}}{r^{2}}-\frac{1}{r^{2}}-k^{2}\right]\delta v_{\theta}+2\eta_{s}\frac{im}{r^{2}}\delta v_{r}\\
0 & = & \left(\partial_{r}+\frac{1}{r}\right)\delta\sigma_{rz}+\frac{im}{r}\delta\sigma_{\theta z}+ik\delta\sigma_{zz}-ik\delta p,\\
 &  & +\eta_{s}\left[\partial_{r}^{2}+\frac{1}{r}\partial_{r}-\frac{m^{2}}{r^{2}}-k^{2}\right]\delta v_{z}
\end{eqnarray*}
and incompressibility states that
\[
0=\left(\partial_{r}+\frac{1}{r}\right)\delta v_{r}+\frac{im}{r}\delta v_{\theta}+ik\delta v_{z}.
\]
 
When the shear-thinning parameter $\alpha$ vanishes, these linear equations
reduce to those derived by Avgousti and Beris \cite{Avgousti1993} for an
Oldroyd-B fluid; as a matter of fact, we have spotted slight differences with
Ref.~\cite{Avgousti1993}, which we believe are typos in that publication.

These equations involve three non-dimensional quantities: the shear-thinning
parameter $\alpha$, the {}``bare Weissenberg'' number $\tau_{\alpha}\dot{\gamma}^{\star}$,
and the relative Newtonian viscosity $\nicefrac{\eta_{s}}{\eta_{0}}$,
on top of the (fixed) relative curvature $\epsilon$.

\subsection{Linear stability analysis: general trends and application to dense
suspensions}

\subsubsection{Consistency of the algorithm for $\alpha\rightarrow0$, $\eta_{s}\rightarrow0$
and influence of the shear rate}

\begin{figure}
\begin{centering}
\subfloat[
$s^{\star}$ \emph{vs. }applied shear rate $\dot{\gamma}$ for vanishing
Newtonian viscosity ($\nicefrac{\eta_{s}}{\eta_{0}}=10^{-6}$)
(UCM model).
]{\begin{centering}
\includegraphics[width=7cm]{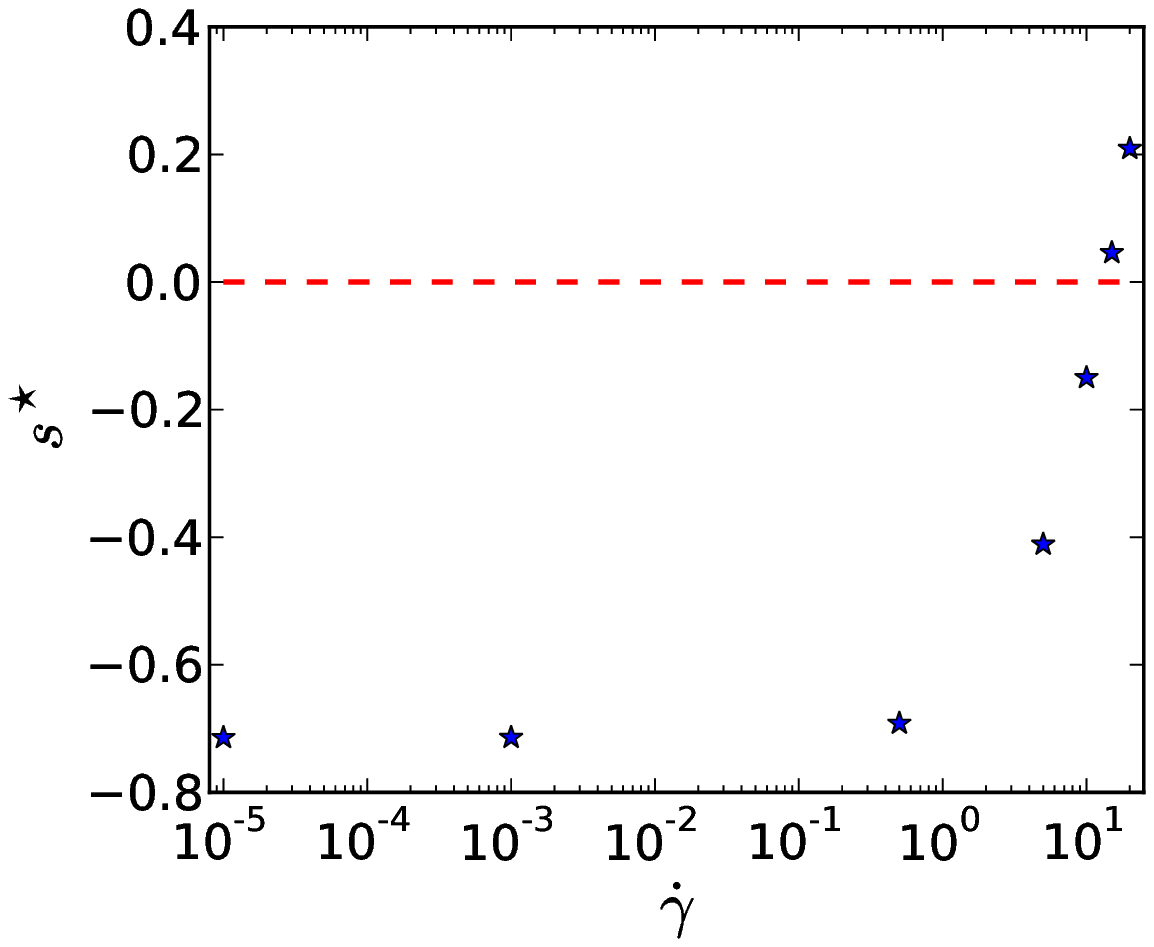}
\par\end{centering}

}\hspace*{0.3cm}\subfloat[\label{fig:PeakVsBeta} $s^{\star}$ \emph{vs.}
 $\eta_s$, at fixed shear rate ($\dot{\gamma}=20$).]
{\includegraphics[width=7cm]{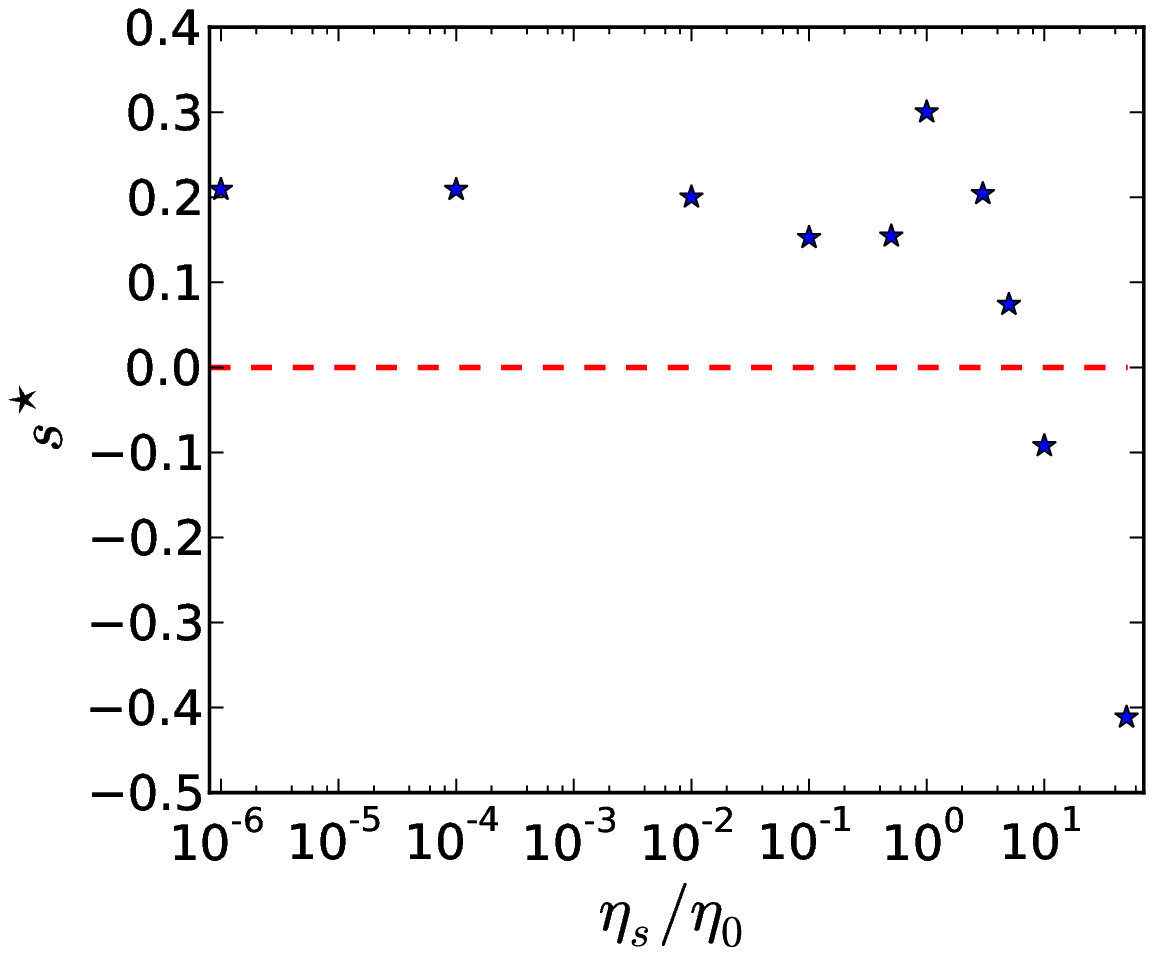}
}
\par\end{centering}

\caption{\label{fig:PeakVsWi}Maximal growth rate $s^{\star}$ (over all modes)
in the limit of a non-shear-thinning fluid, $\alpha=3\cdot10^{-4}$,
$\tau_{\alpha}=1.4$. At $\dot{\gamma}=20$, the most unstable modes
are at $m\simeq1$, $k\simeq70$. The dashed red line separates stable
base flows (below the line) from unstable ones (above).}
\end{figure}

To check the validity of our algorithm, we have first set the shear-thinning
parameter $\alpha$ and the Newtonian viscosity $\eta_{s}$ to values
close to zero. A conventional Upper Convected Maxwell (UCM) model
should then be recovered, in which case, following Ref.~\cite{Larson1990},
some axisymmetric modes (at $m=0$) become unstable at sufficiently
large applied shear rates. This is indeed the case in our simulations,
as soon as the shear rate grows larger than a critical shear rate
comparable to that reported in Ref.~\cite{Larson1990}. However,
we also observe that some non-axisymmetric modes, associated to small
azimuthal wavenumbers $m=\mathcal{O}(1)$, are even more unstable
than the axisymmetric ones, in accordance with the literature \cite{Avgousti1993}.
Figure \ref{fig:PeakVsWi} shows the increase of the growth rate $s^{\text{\ensuremath{\star}}}$
of the most unstable mode with the shear rate $\dot{\gamma}$ measured at the
inner cylinder (or the {}``bare Weissenberg'' number $\tau_{\alpha}\dot{\gamma}$).

\subsubsection{Stabilising effect of the Newtonian viscosity}

\begin{figure}
\begin{centering}
\includegraphics[width=7cm]{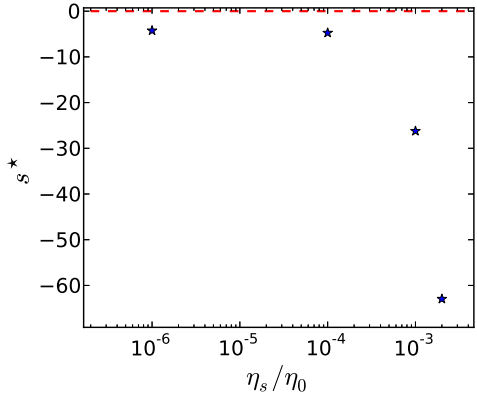}
\par\end{centering}

\caption{\label{fig:s_star_ST_vs_beta}Dependence of the maximal growth rate
$s^{\star}$ on the Newtonian viscosity $\eta_{s}$, for a shear-thinning
fluid with the following parameters: $\alpha=7.7$, $\tau_{\alpha}=1.4$,
$\dot{\gamma}=20$. (The growth rates vary only weakly with $k$ and
$m$ for these parameters; only modes $m\leqslant250$ are considered
here).}

\end{figure}

Including a Newtonian contribution to the stress, \emph{via }a finite
value of the viscosity $\eta_{s}$, tends to stabilise the flow, as
shown in Fig.~\ref{fig:PeakVsBeta}. This stabilising role has already
been reported in the literature on the Oldroyd-B model (\emph{i.e.},
for $\alpha=0$) \cite{Avgousti1993,Shaqfeh1996}; we find that it holds
true for shear-thinning fluids, \emph{i.e.}, when $\alpha$ departs
from zero (see Fig.~\ref{fig:s_star_ST_vs_beta}).

\subsubsection{Stabilisation through shear-thinning\label{sub:Stabilisation-through-shear-thin}}

\begin{figure}
\begin{centering}
\includegraphics[width=7cm]{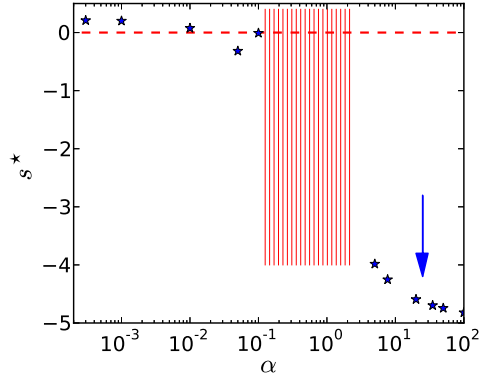}
\par\end{centering}

\caption{\label{fig:s_star_ST_vs_alpha}Dependence of the maximal growth rate
$s^{\star}$ on the shear-thinning parameter $\alpha$, for $\nicefrac{\eta_{s}}{\eta_{0}}=10^{-6}$,
$\tau_{\alpha}=1.4$, $\dot{\gamma}=20$. The arrow points to the
typical values of $\alpha$ of interest here. See the text for the
description of the red vertical bars.}
\end{figure}

\begin{figure}
\begin{centering}
\subfloat[$\alpha=3\cdot10^{-4}$]{\includegraphics[width=7cm]{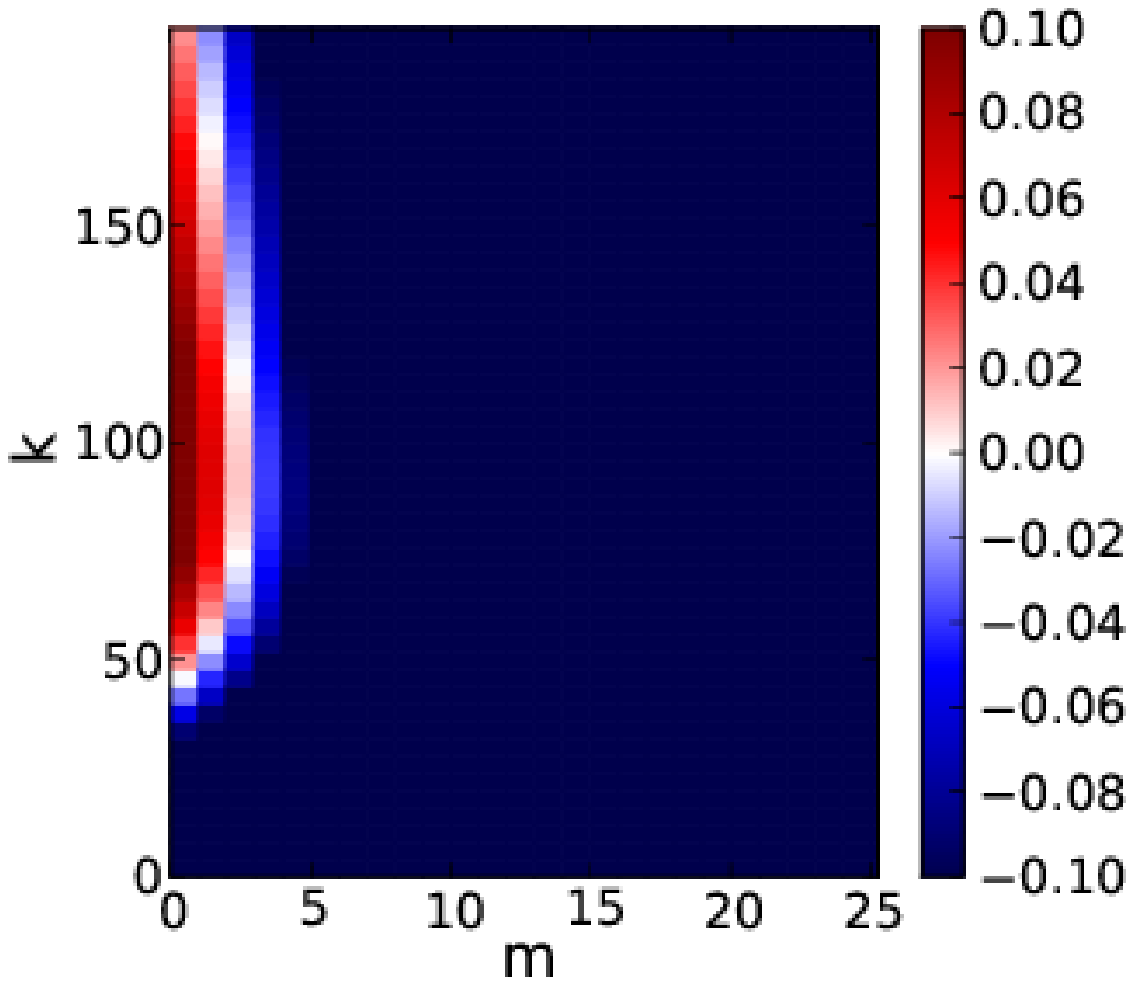}

}\hspace*{0.3cm}\subfloat[$\alpha=10^{-2}$]{\includegraphics[width=7cm]{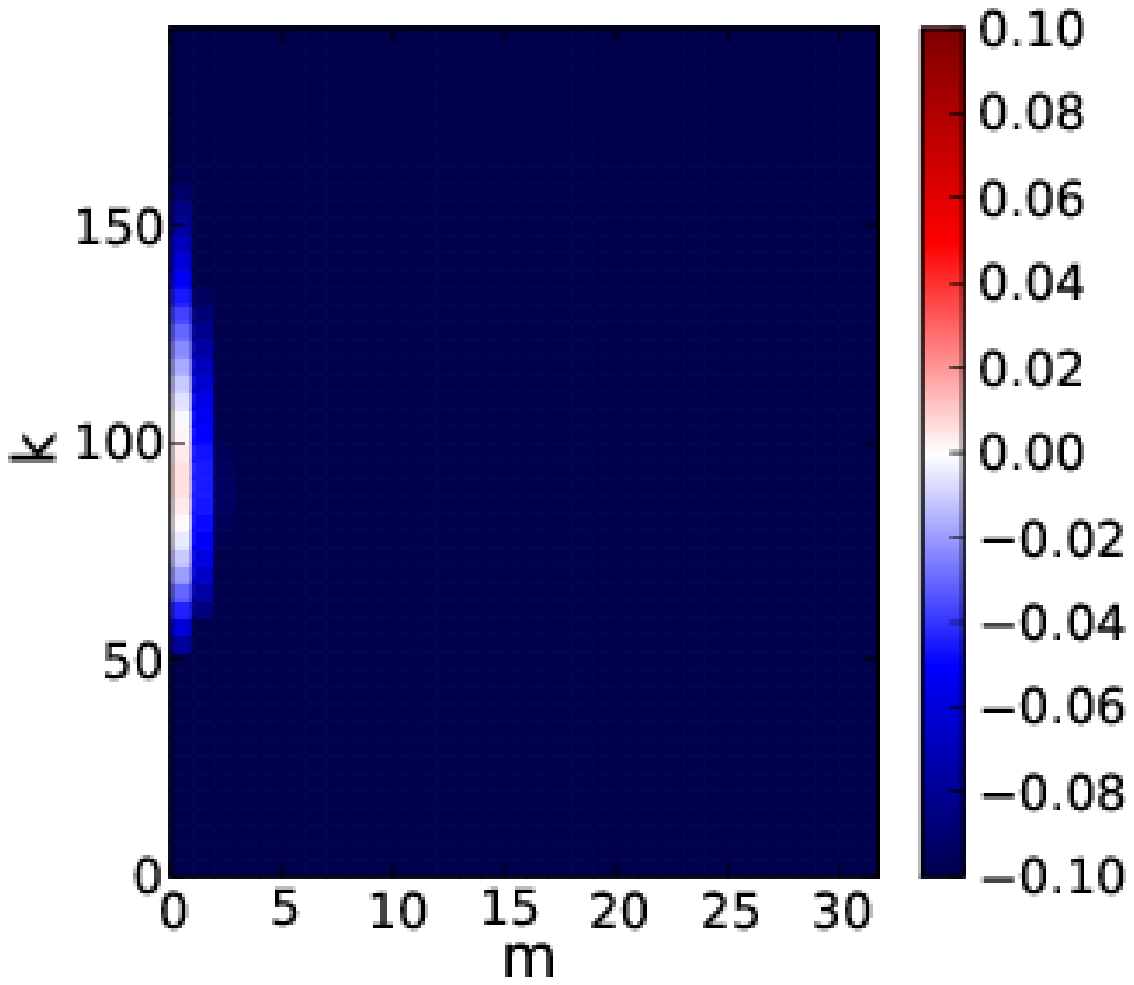}

}
\par\end{centering}

\caption{\label{fig:ColourMaps_MaxGR} Colour maps of the maximal growth rates
associated with each pair of wavenumbers $(m,k)$ for two distinct, but small,
values of $\alpha$ (see captions), at $\dot{\gamma}=20$, with negligible
Newtonian
viscosity ($\nicefrac{\eta_{s}}{\eta_{0}}=10^{-6}$).
Stable modes appear in dark blue.}
\end{figure}

For the values of $\alpha$ of interest here, shear-thinning strongly
suppresses the instability. Figures~\ref{fig:s_star_ST_vs_alpha}-\ref{fig:ColourMaps_MaxGR}
show that, for the model parameters corresponding to the colloidal
suspension at $\phi_{\mathrm{eff}}=0.519$, but with vanishing Newtonian
viscosity, the base flow lies deep within the stable region, whereas,
in the absence of shear-thinning, \emph{i.e.}, were $\alpha$ vanishingly
small, a linear instability would develop, with a peak of instability
around $k\simeq70$, $m\simeq1$.

We should however mention the existence of a limited range of values
of $\alpha$, \emph{outside} the experimentally relevant window for dense
colloidal suspensions, namely, $0.3\lesssim\alpha\lesssim3$ and marked
by red vertical bars on Fig.~\ref{fig:s_star_ST_vs_alpha}, which
display unstable modes associated to \emph{abnormally large growth
rates}. These modes are located in a very different region of the
$(k,m)$-plane, namely $k\approx0$ and $m\gg1$. Although these perturbations
have reasonable shapes in space, it is plausible that they are actually
unphysical, but we do not know whether they are intrinsic in our WM-model
or whether they arise because of artifacts in the (well established) numerical
method. In the following, we concentrate on the experimentally relevant range of
parameters.

\subsubsection{Linear stability analysis of dense colloidal suspensions}

We now specifically consider the model parameters used to fit the
rheology of dense colloidal suspensions, (see Table~\ref{tab:Model-parameters}).

Consistently with the strongly stabilising effect of shear-thinning
described in the previous section (Section~\ref{sub:Stabilisation-through-shear-thin}),
we have numerically ascertained the stability of the base flows corresponding
to the suspension at $\phi=0.519$ for rescaled shear
rates
$\dot{\gamma}=10^{-4},\,10^{-2},\,10^{-1}$,
at $\phi_{ \mathrm{eff}}=0.626$ for
$\dot{\gamma}=5\cdot10^{-6},\,5\cdot10^{-4},\,5\cdot10^{-3},\,10^{-2}$
in the experimental range, as well as that with effective volume fraction
$\phi_{\mathrm{eff}}=0.641$ at $\dot{\gamma}=10^{-8},\,10^{-6},\,10^{-4}$.
The relative contribution of the Newtonian viscosity to the stabilisation
of the flow echoes the contribution of the Newtonian stress to the
total stress: at high densities and low shear rates, the effect of
the Newtonian viscosity is negligible.
Admittedly, the more acute shear-thinning displayed by the WM models fitted to
the densest suspensions results in an enhanced stability compared to the
experimental systems; however the fact that they lie so deep in the stable
region suggests that the materials they model, albeit somewhat less
shear-thinning, are also stable with respect to the considered perturbations.
This is further supported by the observed stability of the system at
$\phi=0.519$, which is both less shear-thinning than the others and very well
described by the WM model.

\subsection{Rationalisation with the Pakdel-McKinley criterion}

Before we conclude, it is interesting to note that the stabilising
effects of shear-thinning and of the Newtonian contribution are in
line with the visco-elastic stability criterion proposed by Pakdel
and McKinley in 1996 \cite{Pakdel1996,Morozov2007} (less
general versions of the criterion can be found in earlier publications
\cite{Larson1990}). On the basis of a dimensional analysis of generic
visco-elastic constitutive equations, these authors introduced a dimensionless
number, written $\mathcal{P}$ here, for inertialess visco-elastic
instabilities in curved geometries in analogy to the classical Taylor
number for inertial instabilities, and propounded the idea that the
base flow is susceptible to an instability if this dimensional number
exceeds a given threshold of order unity. The Pakdel-McKinley number
reads
\[
\mathcal{P}\equiv\frac{l_{p}}{\mathcal{R}}\frac{N_{1}}{\Sigma},
\]
where $l_{p}\equiv v_{\theta}\tau$ is the typical distance travelled
by a material substructure (\emph{e.g.}, a polymer chain) along the
base-flow streamline while relaxing, $\mathcal{R}$ is the radius
of curvature of the streamline, $N_{1}$ is the first normal-stress-difference
and $\Sigma$ is the shear stress. For a UCM model, the ratio $\frac{N_{1}}{\Sigma}$
is simply the Weissenberg number and measures the {}``anisotropy
of the normal forces'', as phrased by Morozov and van Saarloos \cite{Morozov2007}.
But even with the present WM model, which is somewhat more complicated,
the criterion based on $\mathcal{P}$ seems to capture the observed
trends: a Newtonian contribution to the stress increases $\Sigma$
without altering $N_{1}$, thereby reducing $\mathcal{P}$ and stabilising
the flow. On the other hand, an enhanced propensity to shear-thinning
results in a decrease of $N_{1}$ as $\left(1+\alpha\tau_{\alpha}\dot{\gamma}\right)^{-2}$,
while $\Sigma$ only decreases as $\left(1+\alpha\tau_{\alpha}\dot{\gamma}\right)^{-1}$
(for $\eta_{s}\rightarrow0$); as a result, $\mathcal{P}$ is reduced,
which also explains the enhanced stability of the flow. 

The Newtonian stress contribution and the propensity to shear-thinning 
thus act in synergy, but, for the rheology of the very dense colloidal
suspensions of interest here, our results indicate that the impact of
shear-thinning quantitatively prevails over that of the former.

\bigskip{}

\section{Discussion and Outlook}

In conclusion, we have made use of, and partially extended, a general
theoretical framework based on projections onto density modes, to describe
the rheology of dense colloidal suspensions in the vicinity of the
glass transition. Contrary to the previous theoretical works, which focused on
strictly
homogeneous flows, special attention was paid to the fate of spatial
inhomogeneities. However, the intricacy of the formalism forced us
to resort to particularly strong approximations before we could study
the stability of the flow in curved (Taylor-Couette) geometry. At
the expense of these approximations, constitutive equations falling
in the White-Metzner class were obtained; the resulting model was
shown to capture the shear-thinning properties of the material and
to be reasonably consistent with experimental measurements of the
linear rheology and steady-state rheology of the suspensions, although not in a
strictly quantitative way. Eventually, we analysed the stability
of the visco-elastic flow and brought evidence that, in the experimental
range of shear rates, shear-thinning (and to a \emph{much} lesser
extent the Newtonian stress contribution) strongly stabilise the
flow. This may explain why visco-elastic instabilities have been observed
in a variety of visco-elastic fluids, but not in the dense suspensions
under consideration here: the flow strains and destroys the microstructure
of the material, so much so that, if a material volume is displaced
by a perturbation, the memory of the stress that it carries, through
its microstructure, is suppressed by the flow too quickly to allow the
possibility of a feedback mechanism. 

\bigskip{}

\paragraph*{Acknowledgments}
We thank Miriam Siebenb\"urger for providing us with the experimental
rheological data. AN thanks Alexander Morozov for introducing him
to the pseudo-spectral method, Fabian Frahsa for enlightening discussions,
and the DFG-funded research unit FOR 1394 project P3 for partial funding of his stay at the University of Konstanz.

\appendix

\section{Operator formulation of the switch to the auxiliary
frame\label{sec:Operator-formulation-of}}

In Section~\ref{sub:Advection_term_gal}, the switch to the auxiliary
frame at a specific point was presented as a change of coordinates.
However, it may also be conducted with the operator formalism.
Indeed, consider an auxiliary system  (denoted by tildes) with the following
Smoluchowski
operator

\[
\tilde{\Omega}^{\dagger}(\Gamma,t)\equiv\sum_{i=1}^{N}\left[\partial_{i}+\boldsymbol{F_{i}}\left(\Gamma\right)+\boldsymbol{\tilde{v}}\left(\boldsymbol{r_{i}},t\right)\right]\cdot\partial_{i},
\]
where
\begin{equation}
\boldsymbol{\tilde{v}}\left(\boldsymbol{r},t\right)\equiv\boldsymbol{v}^{\mathrm{solv}}\left(\boldsymbol{r},t\right)-\boldsymbol{v}^{\mathrm{solv}}\left(\boldsymbol{r}_{o}(t),t\right).\label{eq:Smoluchowski_tilde}
\end{equation}
As in Section~\ref{sub:Advection_term_gal}, $\boldsymbol{r}_{o}(t)$
is the position (given by Eq.~\ref{eq:change_of_coordinates}) of the
{}``material point'' advected by the solvent
flow field, with position $\boldsymbol{r}_{o}$ at $t_{o}$.

In the auxiliary system, an observable $\tilde{g}(\boldsymbol{r},\Gamma;t)$
evolves as
\begin{eqnarray*}
\partial_{t}\tilde{g}(\boldsymbol{r},\Gamma;t) & = & \tilde{\Omega}^{\dagger}(\Gamma,t)\tilde{g}(\boldsymbol{r},\Gamma;t).
\end{eqnarray*}
Evaluating an observable in this auxiliary system is therefore equivalent to
evaluating it in the auxiliary (denoted by primes) of 
Section~\ref{sub:Advection_term_gal}: $\tilde{g}(\boldsymbol{r},\Gamma;t)$ and
$g^{\prime}(\boldsymbol{r},\Gamma^{\prime};t)\Big|_{\Gamma^{\prime}=\Gamma}$
have the same time derivative
$\tilde{\Omega}^{\dagger}(\Gamma,t)=\Omega^{\dagger\,\prime}(\Gamma^{\prime},
t)\Big|_{\Gamma^{\prime}=\Gamma}$,
 and they coincide at $t_{o}$, so they are equal at all times.
 
The formal solution of Eq.~\ref{eq:Smoluchowski_tilde} is
\begin{eqnarray*}
\tilde{g}(\boldsymbol{r},\Gamma;t) & = & e_{-}^{\int_{0}^{t}\tilde{\Omega}^{\dagger}(\Gamma,s)ds}\tilde{g}(\boldsymbol{r},\Gamma)\\
 & = & e_{-}^{\int_{0}^{t}\tilde{\Omega}^{\dagger}(\Gamma,s)ds}g(\boldsymbol{r},\Gamma).
\end{eqnarray*}
 Using the shorthand $A^{\dagger}(t)$ for $\boldsymbol{v}^{\mathrm{solv}}\left(\boldsymbol{r}_{o}(t),t\right)\cdot\sum_{i=1}^{N}\partial_{i}$,
we observe that $A^{\dagger}(t)$ commutes with all $\Omega^{\dagger}(s)$.
Indeed,
\begin{eqnarray*}
\Omega^{\dagger}(s)A^{\dagger}(t)g & = & \sum_{i=1}^{N}\left[\partial_{i}+\boldsymbol{F_{i}}\left(\Gamma\right)+\boldsymbol{v}^{\mathrm{solv}}\left(\boldsymbol{r_{i}},s\right)\right]\cdot\partial_{i}\left[\boldsymbol{v}^{\mathrm{solv}}\left(\boldsymbol{r}_{o}(t),t\right)\cdot\sum_{j=1}^{N}\partial_{j}g\right]\\
 & = & \boldsymbol{v}^{\mathrm{solv}}\left(\boldsymbol{r}_{o}(t),t\right)\cdot\sum_{i=1}^{N}\sum_{j=1}^{N}\left[\partial_{j}\partial_{i}g+\partial_{j}\left(\boldsymbol{F_{i}}\cdot\partial_{i}g\right)+\partial_{j}\boldsymbol{v}^{\mathrm{solv}}\left(\boldsymbol{r_{i}},s\right)\partial_{i}g\right],\\
 & = & A^{\dagger}(t)\Omega^{\dagger}(s)g,
\end{eqnarray*}
where we have made use of $\partial_{j}\boldsymbol{v}^{\mathrm{solv}}\left(\boldsymbol{r_{j}},s\right)=0$
and $\sum_{i=1}^{N}\boldsymbol{F_{i}}=0$, hence $\sum_{j=1}^{N}\sum_{i=1}^{N}\partial_{j}\boldsymbol{F_{i}}=\boldsymbol{0}$.
It follows that
\begin{eqnarray*}
\tilde{g}(\boldsymbol{r},\Gamma;t) & = & e_{-}^{-\int_{0}^{t}ds\boldsymbol{v}^{\mathrm{solv}}\left(\boldsymbol{r}_{o}(s),s\right)\cdot\sum_{i=1}^{N}\partial_{i}}e_{-}^{\int_{0}^{t}\Omega^{\dagger}(\Gamma,s)ds}g(\boldsymbol{r},\Gamma),\\
 & = & e_{-}^{-\int_{0}^{t}ds\boldsymbol{v}^{\mathrm{solv}}\left(\boldsymbol{r}_{o}(s),s\right)\cdot\sum_{i=1}^{N}\partial_{i}}g(\boldsymbol{r},\Gamma;t)
\end{eqnarray*}
and, if $g$ does not depend intrinsically on space,
\[
\tilde{g}(\boldsymbol{r},\Gamma;t)=e_{-}^{\int_{0}^{t}ds\boldsymbol{v}^{\mathrm{solv}}\left(\boldsymbol{r}_{o}(s),s\right)\cdot\partial_{\boldsymbol{r}}}g(\boldsymbol{r},\Gamma;t).
\]
 In particular, the fluctuation advection term emerges when the equation is
differentiated with respect to time.
\[
\partial_{t}g(\boldsymbol{r},\Gamma;t)=e_{-}^{-\int_{0}^{t}ds\boldsymbol{v}^{\mathrm{solv}}\left(\boldsymbol{r}_{o}(s),s\right)\cdot\partial_{\boldsymbol{r}}}\partial_{t}\tilde{g}(\boldsymbol{r},\Gamma;t)-\boldsymbol{v}^{\mathrm{solv}}\left(\boldsymbol{r}_{o}(t),t\right)\cdot\partial_{\boldsymbol{r}}g(\boldsymbol{r},\Gamma;t).
\]
This equation agrees with Eq.~\ref{eq:dt_gs}.

\end{document}